\documentclass[a4paper]{article}%{ieeeaccess}
\usepackage{cite}
\usepackage{algorithm}
\usepackage{amsmath, amssymb, amsfonts}
\usepackage{algorithmic}
\usepackage[dvipdfmx]{graphicx}
\usepackage{color}
\usepackage{textcomp}
\usepackage{url}
\usepackage{comment}
\usepackage{subcaption}
\newtheorem{theorem}{Theorem}
\definecolor{accessblue}{cmyk}{1,  0.3,  0,  0.2}
\definecolor{greycolor}{cmyk}{0, 0, 0, .8}

\begin{document}
%\history{Date of publication xxxx 00,  0000,  date of current version xxxx 00,  0000.}
%\doi{10.1109/ACCESS.2017.DOI}

%\address[1]{Graduate School of Science and Technology,  Kumamoto University,  Japan}
%\address[2]{Graduate School of Engineering,  Kyoto University,  Kyoto,  Japan}

%\markboth
%{H. Okajima et al \headeretal: From Noise to Knowledge: System Identification with Systematic Polytope Construction via Cyclic Reformulation}
%{H. Okajima et al \headeretal: From Noise to Knowledge: System Identification with Systematic Polytope Construction via Cyclic Reformulation}

%\corresp{Corresponding Author: H. Okajima (e-mail: okajima@cs.kumamoto-u.ac.jp)}

\title{From Noise to Knowledge: System Identification with Systematic Polytope Construction via Cyclic Reformulation}
\author{ Hiroshi Okajima, Shun Shirahama, Tatsunori Hayashi and Nobutomo Matsunaga}

\begin{abstract}
% REVISION: v45 - Abstract rewritten for balance (~210 words, 7 sentences): key concept "intentional periodicity induction" introduced; N-tradeoff compressed; diagnostic caveat shortened; verification details consolidated.
Model-based robust control requires not only accurate nominal models but also systematic uncertainty representations to guarantee stability and performance. However, constructing polytopic uncertainty models typically demands multiple experiments or a priori structural assumptions.
This paper proposes an identification framework based on \emph{intentional periodicity induction}, in which cyclic reformulation with period $N$ is applied to a linear time-invariant system to interpret noise-induced parameter fluctuations as a structured manifestation of estimation uncertainty. The $N$ parameter sets obtained from a single identification experiment---which would coincide in the noise-free case---are used as polytope vertices, providing systematic control over the granularity of the uncertainty description through the choice of $N$. The practical utility of the constructed polytope is demonstrated through robust $H_\infty$ state-feedback synthesis via LMI optimization at the polytope vertices; the synthesis uses only noisy identification data and is shown across Monte Carlo trials to stabilize the true plant with only marginal conservatism. Complementarily, a diagnostic assessment based on the best in-polytope point confirms that the polytope captures meaningful uncertainty information. For a third-order system under Gaussian and uniform noise, a comparison with bootstrap-inspired resampling baselines indicates that cyclic reformulation provides a competitive or favorable trade-off by utilizing the full data record; the construction is further validated on a fourth-order MIMO system.
\end{abstract}

\maketitle

\section{INTRODUCTION}
Model-based control has become essential in control system design, requiring accurate mathematical models that represent the dynamic characteristics of plants. System identification is a fundamental method for estimating models from input-output data\cite{id00,katayama05,n4sid,moesp,verhaegen07,qin06,kawaguchi23}. System identification continues to advance through integration with machine learning techniques, including neural network-based approaches for nonlinear dynamics \cite{forgione21,jacobs}, sparse identification methods for equation discovery \cite{isti}, and data-driven optimization frameworks for probabilistic performance guarantees \cite{pinn_control22}. 

Matrix polytope representation enables the application of powerful design techniques based on linear matrix inequalities (LMIs) \cite{ly00} and has been extensively used in robust control design for discrete-time systems \cite{ly01}. Advanced control methods such as parameter-dependent Lyapunov functions \cite{ly02,oliveira02}, gain-scheduled controllers \cite{apkarian95}, 
multi-objective output-feedback control \cite{scherer97}, and robust filtering techniques \cite{geromel99} all rely on polytopic uncertainty models to ensure guaranteed performance under uncertainty. The effectiveness of polytopic uncertainty modeling has been demonstrated in various applications including model error suppression compensators \cite{s10}.

Despite the widespread use of polytopic models in control design, systematic identification methods that directly derive polytope-type models from experimental data remain underdeveloped. Several approaches have been proposed to address this challenge. 
% REVISION: R1-C2, R2-C1-2 (paragraph reorganization, Bootstrap added, Table 1 removed)
Among these, the approaches most relevant to ours aim to quantify model uncertainty from a single identification experiment, and the idea closest in spirit is found in resampling and subsampling techniques originating from the bootstrap literature in statistics and signal processing\cite{new44,new45}. These techniques generate multiple parameter estimates from a single dataset and naturally yield confidence regions for the identified parameters; although they have not been established as a systematic polytope construction method for control design, the underlying idea of producing multiple estimates from one dataset is closely related to ours, and a data-partitioning variant of this idea serves as the quantitative baseline in Section~\ref{sec:bootstrap}. % REVISION: v33 - Removed forward reference to Section V; v49 - reordered so bootstrap (closest in spirit) precedes set membership, avoiding misleading impression that set membership is the most related
A more distantly related line of work is set membership identification\cite{milanese85,t03,t04,t01,t02,t05}, which constructs parameter sets under bounded noise assumptions by formulating the identification as a constraint satisfaction problem and provides theoretical guarantees for true parameter inclusion within the estimated set; its reliance on a deterministic noise bound, however, places it in a fundamentally different setting from the stochastic noise model adopted in this work. Adaptive identification using multiple models\cite{narendra97} employs switching or blending mechanisms between candidate models, though these approaches may lack systematic uncertainty quantification and require careful model selection strategies. Most of these approaches assume that the polytopic structure is known a priori or rely on heuristic model selection procedures, limiting their practical applicability when the uncertainty structure must be systematically inferred from data.

% REVISION: v33 - Condensed related frameworks paragraph
% REVISION: v42 - scenario references split: calafiore06 is for robust control design, sanchez98 for robust systems theory (Reviewer concern F)
Beyond methods aimed at uncertainty quantification per se, several frameworks employ multiple linear models or polytopic representations in related contexts. Gain scheduling \cite{lawrence95, shamma92} is a control design methodology that switches between controllers designed at different operating points, requiring a priori knowledge of scheduling variables. The scenario approach to robust control design \cite{calafiore06} and the broader robust systems framework \cite{sanchez98} provide probabilistic or worst-case guarantees by sampling or otherwise characterizing the uncertainty set, but they operate at the control-synthesis stage rather than as identification methods and typically assume that the uncertainty description is already available or directly samplable. LPV system identification \cite{toth10, felici07} targets systems whose dynamics genuinely depend on measurable scheduling variables, rather than uncertainty quantification of a single LTI plant. Iterative identification-control co-design \cite{gevers03, bombois06} optimizes experiment design for robust control but requires multiple experiments. While these frameworks share the use of polytopic or multi-model representations, they address fundamentally different problem settings from the one considered here.

Cyclic reformulation\cite{s8,bittanti05} has been proposed as a time-invariant reformulation method for time-varying systems. The cycling approach enables the transformation of periodic time-varying systems into forms that can be handled as time-invariant systems. This transformation is particularly valuable as it preserves the system's essential dynamics while allowing the application of well-established time-invariant identification techniques such as subspace methods\cite{verhaegen92,moesp,qin06,tanaka01,n4sid,oku07}. In our previous work, we have proposed identification methods for periodic time-varying systems using this cycling approach\cite{okajima01} and its application to multi-rate systems\cite{okajima02}. These methods demonstrated that cyclic reformulation can systematically handle time-varying and multi-rate scenarios by converting them into equivalent time-invariant identification problems.

% REVISION: v45 - rephrased the sentence about N. The former version stated that "the single parameter N simultaneously controls both the periodicity structure and the polytope complexity", which suggested two independent attributes being controlled together. In fact, the period N is the only design parameter, and the number of vertices is a direct consequence of that choice. The revised sentence states this one-way relationship plainly.
In this research, we propose intentional periodicity induction for linear time-invariant systems. By deliberately applying cyclic reformulation with 
period $N$ to time-invariant systems, we systematically construct polytopic uncertainty models from noise-induced parameter variations. The period $N$ is the sole design parameter of the construction, and the number of polytope vertices is determined by this choice: applying cyclic reformulation with period $N$ to an LTI system yields exactly $N$ parameter sets by construction.

The key insight is that when a linear time-invariant system is subjected to cyclic reformulation with period $N$, noise effects cause the $N$ resulting 
parameter sets to vary slightly around the true system parameters. Rather than viewing these variations as problematic, we constructively utilize them 
as natural vertices for polytopic uncertainty models. This approach differs fundamentally from existing methods by directly exploiting noise-induced 
parameter variations rather than treating noise as constraints or disturbances to be bounded or eliminated.
% REVISION: R1-C2, R2-C1-2 (Table 1 replaced by inline qualitative positioning)
% REVISION: v33 - Shortened positioning paragraph
% REVISION: v45 - removed the explicit disclaimer about not comparing with set membership identification. The differing noise assumptions are already stated in the related-works paragraph above, so restating them here as a reason for omission was redundant and unnecessarily defensive. The useful information --- that bootstrap-inspired resampling serves as the quantitative baseline --- is retained in a single concise clause.
Compared to these methods, the proposed approach occupies a distinctive position: it operates under standard stochastic noise assumptions, requires only a single identification experiment, and produces a polytopic model whose vertices are directly usable for LMI-based robust control design without a priori structural knowledge of the uncertainty. As a quantitative baseline that shares the stochastic setting and the single-dataset multi-estimate idea, two data-partitioning resampling schemes are evaluated against the proposed method in Section~\ref{numerical_ex}.

The proposed method transforms established cyclic reformulation techniques into a systematic tool for uncertainty quantification, enabling derivation of polytopic models from single identification experiments without requiring a priori structural assumptions about the uncertainty structure. This novel application of cyclic reformulation to linear time-invariant systems represents a significant departure from its traditional use in periodic time-varying system identification.

The main contributions of this work are fourfold: (1) extending cyclic reformulation from its conventional use in periodic time-varying system identification to systematic polytope construction for time-invariant systems, (2) a constructive approach that exploits noise-induced parameter variations as polytope vertices rather than treating noise as disturbances to eliminate, (3) demonstration that the constructed polytope provides meaningful uncertainty information for robust $H_\infty$ control design via LMI-based synthesis (Section~\ref{sec:robust_control}), and (4) numerical validation including comparison with data-partitioning resampling baselines and evaluation across multiple system orders and noise distributions, indicating that the constructed polytope contains in-polytope points closer to the true plant than single-model estimates and these baselines on the tested cases. % REVISION: v42 - bootstrap terminology unified, "consistent improvement" softened (Reviewer concerns B, E)

%Our key innovation is a fundamental reinterpretation of noise in system identification. Traditional approaches treat noise-induced parameter variations as estimation errors that should be minimized. In contrast, we exploit these variations as valuable information about system uncertainty. This paradigm shift from "noise avoidance" to "noise exploitation" enables us to generate robust polytopic models from a single experiment. Rather than fighting noise, we harness it to construct systematic uncertainty representations.

\section{MATRIX POLYTOPE REPRESENTATION OF PLANTS}

\subsection{Plant with Noise}

In this section, we formulate the plant as a linear time-invariant system with noise. The plant $P$ is given as an $n$-th order discrete-time linear time-invariant system as in (\ref{eq:noise_system}).

\begin{eqnarray}
\label{eq:noise_system}
x(k+1) &=& Ax(k) + Bu(k) + d_u(k) \nonumber \\
y(k) &=& Cx(k) + Du(k) + d_y(k)
\end{eqnarray}

Here, $x(k) \in \mathbb{R}^{n \times 1}$ is the state vector, $u(k) \in \mathbb{R}^{m \times 1}$ is the input vector, $y(k) \in \mathbb{R}^{q \times 1}$ is the output vector, $d_u(k) \in \mathbb{R}^{n \times 1}$ is the process noise, and $d_y(k) \in \mathbb{R}^{q \times 1}$ is the observation noise. The system matrices are $A \in \mathbb{R}^{n \times n}$, $B \in \mathbb{R}^{n \times m}$, $C \in \mathbb{R}^{q \times n}$, and $D \in \mathbb{R}^{q \times m}$. In this paper, we assume the pair $(C,A)$ is observable and $(A,B)$ is controllable. 

The process noise $d_u(k)$ and observation noise $d_y(k)$ are assumed to be unknown disturbances that affect the system dynamics and measurements, respectively. These noise components represent modeling uncertainties, external disturbances, and measurement errors that are inevitably present in practical control systems.

\subsection{Matrix Polytope Representation}

Matrix polytope representation is an effective method for expressing system uncertainty by enclosing it with vertex matrices. For a system with polytope-type uncertainty, we consider the discrete-time linear time-invariant system given by (\ref{eq:polytope_system}).

\begin{eqnarray}
\label{eq:polytope_system}
x(k+1) &=& A(\lambda)x(k) + B(\lambda)u(k) \nonumber \\
y(k) &=& C(\lambda)x(k) + D(\lambda)u(k)
\end{eqnarray}

The matrix polytope representation is defined as:

\begin{eqnarray}
\label{eq:polytope_def}
A(\lambda) &:=& \sum_{i=0}^{N-1} \lambda_i A_i, \quad B(\lambda) := \sum_{i=0}^{N-1} \lambda_i B_i \nonumber \\
C(\lambda) &:=& \sum_{i=0}^{N-1} \lambda_i C_i, \quad D(\lambda) := \sum_{i=0}^{N-1} \lambda_i D_i \nonumber \\
\varepsilon &:=& \left\{ \lambda \in \mathbb{R}^N : \lambda_i \geq 0, \sum_{i=0}^{N-1} \lambda_i = 1 \right\}
\end{eqnarray}

Here, $A_i \in \mathbb{R}^{n \times n}$, $B_i \in \mathbb{R}^{n \times m}$, $C_i \in \mathbb{R}^{q \times n}$, and $D_i \in \mathbb{R}^{q \times m}$ are given matrices called vertex matrices. The set $\varepsilon$ is called the standard simplex. It is assumed that the exact value of $\lambda$ cannot be obtained, but the information that $\lambda \in \varepsilon$ is available. The model without noise terms is considered for control design purposes. 

%The polytope representation enables systematic application of powerful control design techniques. Once a polytopic uncertainty model is obtained, various design methods become directly applicable: LMI-based robust controller synthesis \cite{ly00,ly01}, parameter-dependent Lyapunov function approaches \cite{ly02}, gain-scheduled control \cite{apkarian95}, multi-objective output-feedback design \cite{scherer97}, and robust filtering \cite{geromel99}. This established design framework provides strong motivation for systematic polytope identification—deriving polytopic models from experimental data enables immediate application of these well-developed robust control techniques.

The polytope representation enables systematic application of powerful control design techniques. Once a polytopic uncertainty model is obtained, 
various design methods become directly applicable: LMI-based robust controller synthesis \cite{ly00}, stability analysis via common Lyapunov functions \cite{ly01}, 
and parameter-dependent Lyapunov function (PDLF) approaches that significantly reduce conservatism \cite{ly02,oliveira02,ebihara05}. 
Further extensions include gain-scheduled control \cite{apkarian95}, multi-objective output-feedback design \cite{scherer97}, and robust 
filtering \cite{geromel99}. These PDLF-based methods exploit the polytope vertex structure to construct less conservative stability certificates 
and controller designs, demonstrating that the polytope model serves as a versatile foundation for modern robust control synthesis.
% REVISION: v33 - Removed forward-looking "We note that..." paragraph from Sec II-B.
% Polytope containment discussion moved to Sec IV-C/IV-D.

\subsection{Parameter Estimation Error and Polytope Model Hypothesis} \label{model_hypothesis}

In practical system identification under noisy environments, conventional 
methods yield a single parameter set $(A_{\text{conv}}, B_{\text{conv}}, 
C_{\text{conv}}, D_{\text{conv}})$ that deviates from the true system 
$(A, B, C, D)$ due to process noise $d_u(k)$ and observation noise $d_y(k)$. 
The magnitude of these estimation errors depends on noise characteristics, 
data length, and the identification algorithm employed.

We propose an alternative approach: construct a polytope model from multiple 
parameter sets $\{(A_{mi}, B_{mi}, C_{mi}, D_{mi})\}_{i=0}^{N-1}$ obtained 
through systematic identification, and regard any convex combination
\begin{eqnarray}
\label{eq:polytope_model}
A(\lambda) &=& \sum_{i=0}^{N-1} \lambda_i A_{mi}, \quad
B(\lambda) = \sum_{i=0}^{N-1} \lambda_i B_{mi} \nonumber \\
C(\lambda) &=& \sum_{i=0}^{N-1} \lambda_i C_{mi}, \quad
D(\lambda) = \sum_{i=0}^{N-1} \lambda_i D_{mi}
\end{eqnarray}
with $\lambda \in \varepsilon$ as a candidate in-polytope model. Unlike the single-point estimate produced by conventional identification, the proposed method yields an entire set of models---the polytope itself---rather than a single point. % REVISION: v42 - rewrote to emphasize polytope-as-set, moved strong hypothesis to Section V

% REVISION: v44 - Working Hypothesis and Remark 1 relocated to Section IV-F, where the concepts required to state them (cyclic reformulation, N vertices, phase-dependent noise realizations) and the diagnostic tool used to test them (the best in-polytope point lambda*) appear together. Only a brief pointer is retained here so that Section II-C remains focused on the single-point vs. set-of-models contrast.
The motivation for constructing such a set rather than a single point, the underlying mechanism, and the precise hypothesis that is tested numerically are stated in Section~\ref{sec:hypothesis_diagnosis}, once the required concepts (cyclic reformulation, $N$ vertices, phase-dependent noise realizations) and the identification algorithm have been introduced.

\section{CYCLIC REFORMULATION FOR PERIODIC TIME-VARYING SYSTEMS}\label{sectioniii}

Cyclic reformulation\cite{s8} transforms periodic time-varying systems with 
period $N$ into equivalent time-invariant forms, enabling application of 
standard identification methods\cite{moesp}. Our previous work developed 
comprehensive identification frameworks for periodic systems\cite{okajima01} 
and extended this approach to multirate sensing environments\cite{okajima02}, 
demonstrating the versatility and broad applicability of cyclic reformulation.

The present work applies this framework in a fundamentally different direction. 
Rather than identifying periodic time-varying systems (the conventional use), 
we apply cyclic reformulation with period $N$ to linear time-invariant systems 
for systematic polytope construction (Section~\ref{seciv}). Noise effects 
naturally generate $N$ parameter sets that vary around true system parameters, 
enabling polytope construction from a single identification experiment.

The following subsections review the cyclic reformulation framework, providing 
the foundation for this novel application.

\subsection{System Formulation and Cyclic Structure}

Consider a linear periodic time-varying system with period $N$:

\begin{eqnarray}
\label{eq:periodic_system}
x(k+1) &=& A(k)x(k) + B(k)u(k) + d_u(k) \nonumber \\
y(k) &=& C(k)x(k) + D(k)u(k) + d_y(k)
\end{eqnarray}

where the system matrices are periodic with period $N$:
\begin{eqnarray}
\label{eq:periodicity}
A(k+N) &=& A(k), \quad B(k+N) = B(k) \nonumber \\
C(k+N) &=& C(k), \quad D(k+N) = D(k)
\end{eqnarray}

Within each period, the system matrices can be expressed as:
\begin{eqnarray}
\label{eq:period_matrices}
A(k) &=& A_i \text{ with } i := k \mod N \nonumber \\
B(k) &=& B_i \text{ with } i := k \mod N \nonumber \\
C(k) &=& C_i \text{ with } i := k \mod N \nonumber \\
D(k) &=& D_i \text{ with } i := k \mod N \nonumber
\end{eqnarray}

This formulation captures the essential characteristic of periodic time-varying systems where the dynamics change cyclically over time with a known period $N$. Bittanti and Colaneri \cite{s8} established the theoretical foundation for cyclic reformulation of discrete-time periodic systems through invariant representations, demonstrating how periodic time-varying dynamics can be systematically transformed into time-invariant forms.

For a given period $N$, the cycled input signal $\check{u}(k) \in \mathbb{R}^{Nm}$ is constructed from the original input $u(k) \in \mathbb{R}^{m}$ as follows:

\begin{eqnarray}
\label{eq:cycled_input}
\check{u}(0) &=& \begin{bmatrix} u(0) \\ O_{m,1} \\ \vdots \\ O_{m,1} \end{bmatrix}, \quad 
\check{u}(1) = \begin{bmatrix} O_{m,1} \\ u(1) \\ \vdots \\ O_{m,1} \end{bmatrix}, \nonumber \\
&\cdots&, \quad \check{u}(N-1) = \begin{bmatrix} O_{m,1} \\ \vdots \\ O_{m,1} \\ u(N-1) \end{bmatrix}
\end{eqnarray}

The cycled input has a unique non-zero sub-vector at each time step, with the non-zero element cyclically shifting through the vector positions. Similarly, the cycled output $\check{y}(k) \in \mathbb{R}^{Nq}$ and noise signals $\check{d}_u(k), \check{d}_y(k)$ are constructed in the same manner.

The cyclic reformulation of the periodic time-varying system (\ref{eq:periodic_system}) with period $N$ results in the following time-invariant representation:

\begin{eqnarray}
\label{eq:cyclic_system}
\check{x}(k+1) &=& \check{A}\check{x}(k) + \check{B}\check{u}(k) + \check{d}_u(k) \nonumber \\
\check{y}(k) &=& \check{C}\check{x}(k) + \check{D}\check{u}(k) + \check{d}_y(k)
\end{eqnarray}

where the system matrices have the following cyclic structure for periodic time-varying systems:

\begin{eqnarray}
\label{eq:cyclic_matrices}
\check{A} &=& \begin{bmatrix}
O_{n,n} & O_{n,n} & \cdots & O_{n,n} & A_{N-1} \\
A_0 & O_{n,n} & \cdots & O_{n,n} & O_{n,n} \\
O_{n,n} & A_1 & \cdots & O_{n,n} & O_{n,n} \\
\vdots & \vdots & \ddots & \vdots & \vdots \\
O_{n,n} & O_{n,n} & \cdots & A_{N-2} & O_{n,n}
\end{bmatrix} \nonumber \\
\check{B} &=& \begin{bmatrix}
O_{n,m} & O_{n,m} & \cdots & O_{n,m} & B_{N-1} \\
B_0 & O_{n,m} & \cdots & O_{n,m} & O_{n,m} \\
O_{n,m} & B_1 & \cdots & O_{n,m} & O_{n,m} \\
\vdots & \vdots & \ddots & \vdots & \vdots \\
O_{n,m} & O_{n,m} & \cdots & B_{N-2} & O_{n,m}
\end{bmatrix}
\end{eqnarray}

\begin{eqnarray}
\label{eq:cyclic_output_matrices}
\check{C} &=& \text{diag}(C_0, C_1, \ldots, C_{N-1}) \nonumber \\
\check{D} &=& \text{diag}(D_0, D_1, \ldots, D_{N-1})
\end{eqnarray}

The dimensions of the cyclic system are: $\check{A} \in \mathbb{R}^{Nn \times Nn}$, $\check{B} \in \mathbb{R}^{Nn \times Nm}$, $\check{C} \in \mathbb{R}^{Nq \times Nn}$, and $\check{D} \in \mathbb{R}^{Nq \times Nm}$.

\textbf{Cyclic Structure and Properties:} 
The cyclic reformulation transforms an $N$-periodic time-varying system into a time-invariant representation where system matrices $\check{A}, \check{B}, \check{C}, \check{D}$ exhibit characteristic cyclic and block-diagonal structures. Specifically, matrices $\check{A}$ and $\check{B}$ possess cyclic matrix structures as shown in (\ref{eq:cyclic_matrices}), while $\check{C}$ and $\check{D}$ are block diagonal matrices as shown in (\ref{eq:cyclic_output_matrices}). This structural property is fundamental to the identification process, as it enables the systematic extraction of time-varying parameters from the time-invariant representation through coordinate transformation and parameter extraction procedures described in the following subsections.

\textbf{Markov Parameter Sparsity Properties:} Critical sparsity properties of Markov parameters under cyclic reformulation were rigorously analyzed. 
These sparsity properties enable systematic extraction of time-varying 
parameters from identified models. The cyclic shift matrix for dimension 
parameter $d$ is defined as an $Nd \times Nd$ matrix:
\begin{eqnarray}\label{eq:cyclic_shift}
\check S_d =  \left[\begin{array}{ccccc}
          O_{d,d} & I_d & O_{d,d} & \cdots & O_{d,d}\\
          O_{d,d} & O_{d,d} & I_d & \ddots & \vdots\\
          \vdots & \ddots & \ddots & \ddots & O_{d,d}\\
          O_{d,d} & \ddots & \ddots & \ddots & I_d\\
          I_d & O_{d,d} & \cdots & \cdots & O_{d,d} 
     \end{array} \right] \in \mathbb{R}^{Nd \times Nd}
\end{eqnarray}
where $d$ takes values such as $q$ (output dimension), $m$ (input 
dimension), or other structural parameters as needed. $I_d \in \mathbb{R}^{d\times d}$ denotes the identity matrix. 

The Markov parameters of the cycled system are defined as $\check{H}^{(0)} = \check{D}$ and $\check{H}^{(i)} = \check{C}\check{A}^{i-1}\check{B}$ for $i \geq 1$. It was proven \cite{okajima01} that matrices of the form $\check{S}_q^i \check{H}(i+j)\check{S}_m^j$ exhibit block-diagonal structure for any $i, j = 0, 1, \ldots$, where $\check{H}(i)$ represents the Markov parameters of the cycled system. These sparsity properties enable systematic extraction of time-varying parameters from identified models.

\subsection{Subspace Identification with Cycled Signals}

The subspace identification method is applied to the cycled input-output signals $\{\check{u}(k), \check{y}(k)\}$ to obtain the system matrices. This approach leverages the time-invariant nature of the cyclic reformulation while preserving the essential dynamics of the original periodic time-varying system.

The subspace identification procedure yields estimated matrices $(\check{A}_*, \check{B}_*, \check{C}_*, \check{D}_*)$ with dimensions corresponding to the cyclic system. However, these identified matrices typically do not exhibit the desired cyclic structure due to estimation errors and coordinate transformation freedom inherent in subspace methods.

\subsection{Coordinate Transformation for Cyclic Structure Recovery}

To recover the cyclic structure from the identified matrices, a coordinate transformation is applied using a transformation matrix $T \in \mathbb{R}^{Nn \times Nn}$:

\begin{eqnarray}
\label{eq:coordinate_transform}
\check{A}_m &=& T^{-1}\check{A}_* T \nonumber \\
\check{B}_m &=& T^{-1}\check{B}_* \nonumber \\
\check{C}_m &=& \check{C}_* T \nonumber \\
\check{D}_m &=& \check{D}_* 
\end{eqnarray}

The transformation matrix $T$ is constructed to enforce the cyclic structure. Based on the observability properties of the system, $T^{-1}$ is defined as:

\begin{eqnarray}
\label{eq:transformation_matrix}
T^{-1} = \sum_{j=0}^{n-1} \check{F}_j \check{S}_q^{j} \check{C}_* \check{A}_*^{j}
\end{eqnarray}

where $\check{F}_j$ are appropriately chosen block diagonal matrices and $\check{S}_q$ is a cyclic shift matrix (\ref{eq:cyclic_shift}) with $d = q$. 
The following theorem, presented in \cite{okajima01}, provides the theoretical foundation by establishing that controllability and observability are sufficient conditions for the coordinate transformation to provably recover the cyclic reformulation structure, thereby enabling systematic extraction of individual time-varying matrices $A_k, B_k, C_k, D_k$ for $k = 0, \ldots, N-1$.

\begin{theorem}
\label{theorem:cyclic_structure}
Assume that the parameters $(\check{A}_*, \check{B}_*, \check{C}_*, \check{D}_*)$ are obtained via subspace identification based on cycled signals, and that the pairs $(\check{A}_*, \check{B}_*)$ and $(\check{C}_*, \check{A}_*)$ are controllable and observable, respectively. Then, the system matrices $(\check{A}_m, \check{B}_m, \check{C}_m, \check{D}_m)$ obtained by the coordinate transformation (\ref{eq:coordinate_transform}) using the transformation matrix $T$ from (\ref{eq:transformation_matrix}) exhibit the cyclic reformulation structure. 
\end{theorem}

%\begin{proof}
%The proof follows from the block diagonal and cyclic properties of the Markov parameters under the cyclic reformulation. The transformation matrix $T^{-1}$ is constructed such that:
%\begin{enumerate}
%\item $\check{D}_m = \check{D}_*$ maintains block diagonal structure
%\item $\check{B}_m = T^{-1}\check{B}_*$ becomes a cyclic matrix
%\item $\check{C}_m = \check{C}_*T$ becomes a block diagonal matrix  
%\item $\check{A}_m = T^{-1}\check{A}_*T$ becomes a cyclic matrix
%\end{enumerate}
%The detailed proof utilizes the properties of cyclic shift matrices and block diagonal structures inherent in the cyclic reformulation framework.
%\end{proof}

Theorem \ref{theorem:cyclic_structure} is derived based on the Markov parameter sparsity Property. The proof relies on the observability Gramian properties under cyclic reformulation, which ensures that $T^{-1}$ defined in (\ref{eq:transformation_matrix}) 
has full rank. For detailed proof, see \cite{okajima01}. %The transformation matrix $T^{-1}$ defined in (\ref{eq:transformation_matrix}) is constructed based on the observability Gramian of the cycled system. 
The key insight is that the matrices $\check{F}_j$ are chosen such that: 1) the product $\check{F}_j \check{S}_q^{j} \check C_* \check A_*^{j}$ extracts specific structural information from the identified system, 2) the sum over $j = 0, \ldots, n-1$ ensures that all observable modes are captured, and 3) the resulting $T^{-1}$ in (\ref{eq:transformation_matrix}) has full rank $Nn$ under the observability assumption. This construction ensures that the transformed system $(\check{A}_m, \check{B}_m, \check{C}_m, \check{D}_m)$ exhibits the desired cyclic structure, enabling the extraction of individual time-varying parameters.

Alternatively, a dual approach based on controllability properties has been proposed in \cite{okajima02}. Instead of computing $T^{-1}$ from observability as in (\ref{eq:transformation_matrix}), the transformation matrix $T$ can be directly constructed using the controllability Gramian:

\begin{eqnarray}
\label{eq:transformation_matrix_dual}
T = \sum_{j=0}^{n-1}  \check{A}_*^{j} \check{B}_* \check{S}_m^{j+1} \check{G}_j 
\end{eqnarray}

where $\check{G}_j$ are appropriately chosen block diagonal matrices and $\check{S}_m$ is a cyclic shift matrix (\ref{eq:cyclic_shift}) corresponding to the input space structure. This dual formulation exploits the controllability properties of the cycled system $(\check{A}_*, \check{B}_*)$, whereas the previous formulation (\ref{eq:transformation_matrix}) exploits the observability properties of $(\check{C}_*, \check{A}_*)$. Explicit constructions of $\check{F}_j$ and $\check{G}_j$ in terms of the observability and controllability structures of the cycled system are given in our previous works~\cite{okajima01,okajima02}; we omit them here for brevity, since they are not used directly in the novel polytope construction developed in this paper.

The two approaches are theoretically equivalent under the assumption that both controllability and observability conditions are satisfied, as stated in Theorem \ref{theorem:cyclic_structure}. The choice between (\ref{eq:transformation_matrix}) and (\ref{eq:transformation_matrix_dual}) depends on the numerical properties of the identified system and computational considerations. In practice, the observability-based approach (\ref{eq:transformation_matrix}) is often preferred when output measurements are more reliable, while the controllability-based approach (\ref{eq:transformation_matrix_dual}) may be advantageous in input-dominant scenarios or multi-rate sampling environments as discussed in \cite{okajima02}.

\subsection{Parameter Extraction for Periodic Time-Varying Systems}

After the coordinate transformation (\ref{eq:coordinate_transform}), the resulting matrices $(\check{A}_m, \check{B}_m, \check{C}_m, \check{D}_m)$ exhibit the desired cyclic structure corresponding to the periodic time-varying system. From these matrices, $N$ different parameter sets can be extracted:

\begin{eqnarray}
\label{eq:parameter_extraction}
\{(A_{mi}, B_{mi}, C_{mi}, D_{mi}) \mid i = 0, 1, \ldots, N-1\}
\end{eqnarray}

These parameter sets correspond to the system matrices at different time instants within the period, representing the time-varying characteristics of the original periodic system. Each parameter set $(A_{mi}, B_{mi}, C_{mi}, D_{mi})$ captures the system dynamics at time instant $i$ within the period $N$, thereby providing a complete characterization of the periodic time-varying behavior.

For periodic time-varying systems, this extraction process successfully recovers the time-varying parameters that describe the system's evolution over each period. The cyclic reformulation framework thus enables the application of time-invariant identification techniques to periodic time-varying systems while preserving all essential dynamic information.

\section{INTENTIONAL PERIODICITY INDUCTION FOR UNCERTAINTY QUANTIFICATION} \label{seciv}

\subsection{Systematic Uncertainty Modeling via Cyclic Reformulation}

%Traditional cyclic reformulation has been developed as a method to transform periodic time-varying systems into time-invariant forms as demonstrated in Section \ref{sectioniii}. In this research, we propose a novel approach by applying this technique in a \textbf{reverse manner} to linear time-invariant systems, enabling systematic construction of polytope-type uncertainty models.
Traditional cyclic reformulation transforms periodic time-varying systems into time-invariant forms (Section \ref{sectioniii}). In this research, we extend this framework: deliberately applying cyclic reformulation to linear time-invariant systems to systematically capture noise-induced uncertainty. We term this approach intentional periodicity induction. 

The key insight is recognizing that when a linear time-invariant system is subjected to cyclic reformulation with period $N$, two distinct situations arise depending on the presence of noise:

\textbf{Ideal situation (noise-free)}: All $N$ parameter sets extracted from the cyclic structure become identical, reflecting the true time-invariant nature of the system.

\textbf{Practical situation (noisy)}: $N$ parameter sets exhibit variations around the true values due to the influence of process and observation noise.

We constructively utilize these noise-induced variations as vertices of a polytopic uncertainty model, rather than treating them as undesirable disturbances to be eliminated. This approach differs from traditional methods that view noise as constraints to be bounded, instead transforming noise-induced variations into systematic uncertainty information for robust model construction.

\subsection{Application to Linear Time-Invariant Systems}

Consider the linear time-invariant system from (\ref{eq:noise_system}). When we apply cyclic reformulation with an arbitrary period $N$ to this system, we obtain:

\begin{eqnarray}
\label{eq:lti_cyclic_system}
\check{x}(k+1) &=& \check{A}_{LTI}\check{x}(k) + \check{B}_{LTI}\check{u}(k) + \check{d}_u(k) \nonumber \\
\check{y}(k) &=& \check{C}_{LTI}\check{x}(k) + \check{D}_{LTI}\check{u}(k) + \check{d}_y(k)
\end{eqnarray}

In the ideal noise-free case, the cyclic matrices would have the following structure:

\begin{eqnarray}
\label{eq:ideal_lti_cyclic_matrices}
\check{A}_{LTI} &=& \begin{bmatrix}
O_{n,n} & O_{n,n} & \cdots & O_{n,n} & A \\
A & O_{n,n} & \cdots & O_{n,n} & O_{n,n} \\
O_{n,n} & A & \cdots & O_{n,n} & O_{n,n} \\
\vdots & \vdots & \ddots & \vdots & \vdots \\
O_{n,n} & O_{n,n} & \cdots & A & O_{n,n}
\end{bmatrix} \nonumber \\
\check{B}_{LTI} &=& \begin{bmatrix}
O_{n,m} & O_{n,m} & \cdots & O_{n,m} & B \\
B & O_{n,m} & \cdots & O_{n,m} & O_{n,m} \\
O_{n,m} & B & \cdots & O_{n,m} & O_{n,m} \\
\vdots & \vdots & \ddots & \vdots & \vdots \\
O_{n,m} & O_{n,m} & \cdots & B & O_{n,m}
\end{bmatrix}
\end{eqnarray}

\begin{eqnarray}
\label{eq:ideal_lti_cyclic_output_matrices}
\check{C}_{LTI} &=& \text{diag}(C, C, \ldots, C) \nonumber \\
\check{D}_{LTI} &=& \text{diag}(D, D, \ldots, D)
\end{eqnarray}

However, in practical noisy environments, the matrices obtained through identification exhibit the following structure after coordinate transformation:

\begin{eqnarray}
\label{eq:actual_cyclic_structure}
\check{A}_m &=& \begin{bmatrix}
O_{n,n} & \cdots & \cdots   & O_{n,n} & A_{m(N-1)}\\
A_{m0} & O_{n,n} & \cdots & \cdots & O_{n,n} \\
O_{n,n} & A_{m1} & \ddots &\ddots & \vdots \\
\vdots & \ddots & \ddots &\ddots & \vdots \\
O_{n,n} & \cdots & O_{n,n} & A_{m(N-2)} & O_{n,n}
\end{bmatrix} \nonumber \\
\check{B}_m &=& \begin{bmatrix}
O_{n,m} & \cdots & \cdots   & O_{n,m} & B_{m(N-1)}\\
B_{m0} & O_{n,m} & \cdots & \cdots & O_{n,m} \\
O_{n,m} & B_{m1} & \ddots &\ddots & \vdots \\
\vdots & \ddots & \ddots &\ddots & \vdots \\
O_{n,m} & \cdots & O_{n,m} & B_{m(N-2)} & O_{n,m}
\end{bmatrix} \nonumber \\
\check{C}_m &=& \text{diag}(C_{m0}, C_{m1}, \ldots, C_{m(N-1)}) \nonumber \\
\check{D}_m &=& \text{diag}(D_{m0}, D_{m1}, \ldots, D_{m(N-1)}) \nonumber 
\end{eqnarray}

where $A_{mi}, B_{mi}, C_{mi}, D_{mi}$ $(i=0,\ldots,N-1)$ are parameters that deviate slightly from the true values $(A, B, C, D)$ due to noise effects.

\subsection{Coordinate Transformation and Parameter Extraction}

When applying the coordinate transformation framework established in Section \ref{sectioniii} to linear time-invariant systems, subspace identification is applied to the cyclic input-output signals $\{\check{u}(k), \check{y}(k)\}$ to obtain system matrices $(\check{A}_*, \check{B}_*, \check{C}_*, \check{D}_*)$. These matrices generally do not exhibit the ideal cyclic structure due to estimation errors and coordinate transformation ambiguities inherent in subspace methods.

The same coordinate transformation procedure as described in Section \ref{sectioniii} 
is applied using a transformation matrix $T \in \mathbb{R}^{Nn \times Nn}$ as in 
(\ref{eq:coordinate_transform}). From the coordinate-transformed system matrices $(\check{A}_m, \check{B}_m, \check{C}_m, \check{D}_m)$, the following $N$ parameter sets are extracted:

\begin{eqnarray}
\label{eq:parameter_sets}
\{(A_{mi}, B_{mi}, C_{mi}, D_{mi}) \mid i = 0, 1, \ldots, N-1\}
\end{eqnarray}

\textbf{Critical Difference from Periodic Time-Varying Case}: In the ideal noise-free case, all $N$ parameter sets would be identical: $(A_{mi}, B_{mi}, C_{mi}, D_{mi}) = (A, B, C, D)$ for all $i = 0, \ldots, N-1$, reflecting the time-invariant nature of the true system. However, in the presence of noise, these parameter sets exhibit variations around the true values: $(A_{mi}, B_{mi}, C_{mi}, D_{mi}) \approx (A, B, C, D) + \delta_i$, where $\delta_i$ represents the noise-induced variation component that differs for each $i$.

% REVISION: v45 - The caveat that "the resulting polytope does not necessarily enclose the true parameters" is deleted from this position. The same caveat is now stated more precisely in Section IV-F (Remark 1 and Observation (i)) as part of the hypothesis-and-diagnosis discussion, which is its logically correct location. Retaining it here --- before the polytope has even been constructed in the next subsection --- interrupted the technical narrative of parameter extraction.
This deviation from the ideal identical parameter sets provides the foundation for systematic polytope construction.

\subsection{Polytope Construction and Theoretical Analysis}

Using the extracted $N$ parameter sets as vertices, we construct a polytope-type uncertainty model:

%\begin{eqnarray}
%\label{eq:polytope_construction}
%A(\lambda) &=& \sum_{i=0}^{N-1} \lambda_i A_{mi} \\
%B(\lambda) &=& \sum_{i=0}^{N-1} \lambda_i B_{mi} \\
%C(\lambda) &=& \sum_{i=0}^{N-1} \lambda_i C_{mi} \\
%D(\lambda) &=& \sum_{i=0}^{N-1} \lambda_i D_{mi}
%\end{eqnarray}
\begin{align}
\label{eq:polytope_construction}
A(\lambda) &= \sum_{i=0}^{N-1} \lambda_i A_{mi}, \quad 
B(\lambda) = \sum_{i=0}^{N-1} \lambda_i B_{mi} \nonumber \\
C(\lambda) &= \sum_{i=0}^{N-1} \lambda_i C_{mi}, \quad 
D(\lambda) = \sum_{i=0}^{N-1} \lambda_i D_{mi}
\end{align}

where $\lambda \in \varepsilon$.

\begin{figure}[t]
\centering
\includegraphics[width=0.85\textwidth]{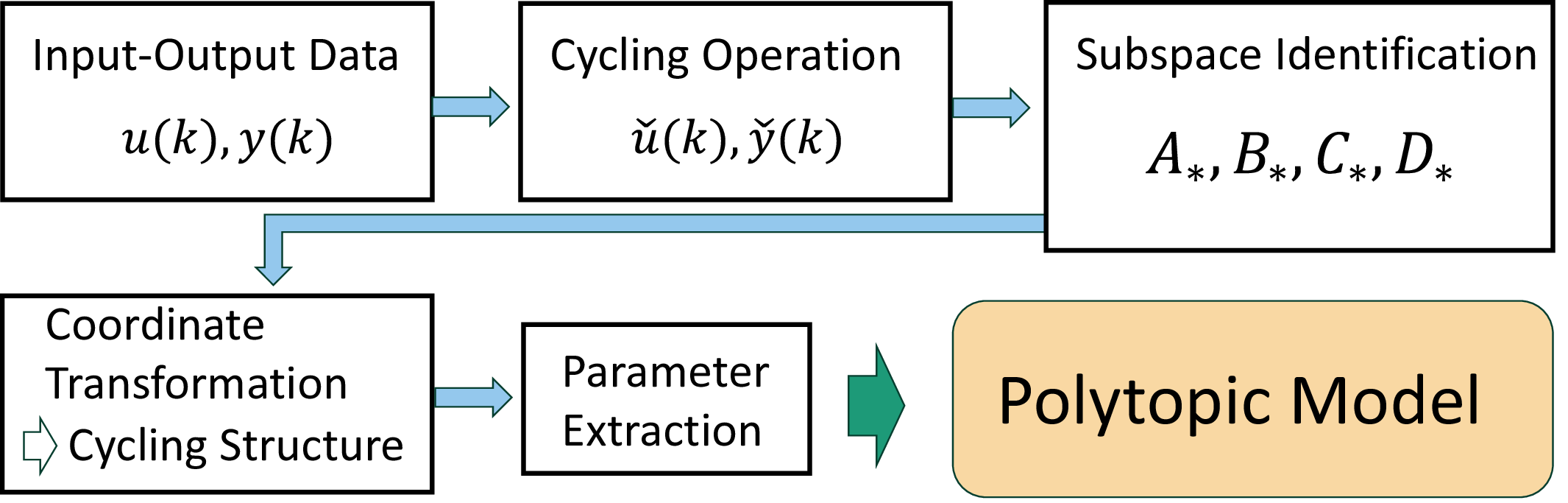}
\caption{Schematic overview of the proposed polytope construction method via intentional periodicity induction.}
\label{fig:workflow}
\end{figure}

The proposed method enables systematic generation of $N$ vertices from a 
single identification experiment, constructively utilizing noise-induced 
variations rather than treating noise as disturbances to eliminate. This 
single-experiment approach reduces experimental cost while providing 
systematic uncertainty quantification suitable for robust control design.
The complete workflow of the proposed method is illustrated in Fig.~\ref{fig:workflow}.

The period $N$ is a designer-selectable parameter that controls the number 
of polytope vertices. Larger $N$ provides more detailed uncertainty representation but increases computational cost as $O((Nn)^3)$ and requires 
sufficient data length. The trade-offs between $N$, data requirements, and 
representation quality are examined through numerical examples in Section~\ref{numerical_ex}.

\subsection{Algorithm}\label{complete_algorithm}

The proposed algorithm takes input-output data $\{u(k), y(k)\}$ for $k=1,\ldots,N_{\text{data}}$ and period setting $N$ as inputs, and outputs an $N$-vertex polytope-type uncertainty model through the following steps:

\begin{enumerate}
\item \textbf{Signal Transformation}: Convert input-output data to cyclic signals $\{\check{u}(k), \check{y}(k)\}$
\item \textbf{Subspace Identification}: Apply subspace identification to cyclic signals to obtain $(\check{A}_*, \check{B}_*, \check{C}_*, \check{D}_*)$
\item \textbf{Coordinate Transformation}: Recover cyclic structure using transformation matrix $T$
\item \textbf{Parameter Extraction}: Extract $N$ parameter sets $\{(A_{mi}, B_{mi}, C_{mi}, D_{mi})\}_{i=0}^{N-1}$ from cyclic matrices
\item \textbf{Polytope Construction}: Construct polytope-type uncertainty model using $N$ sets as vertices
\end{enumerate}

\textbf{Computational Complexity}: Each step contributes to the overall computational cost as follows: signal transformation requires $O(N \cdot N_{\text{data}})$, subspace identification $O((Nn)^3)$, coordinate transformation $O((Nn)^3)$, and parameter extraction $O(N \cdot n^2)$. The overall complexity is dominated by the $O((Nn)^3)$ terms from subspace identification and coordinate transformation. Since the augmented system has dimension $Nn$, larger period $N$ significantly increases computational burden. The selection of $N$ involves a trade-off between computational efficiency and polytope representation quality, with the feasible range depending on available data length $N_{\text{data}}$ and system order $n$ (detailed analysis in Section~\ref{numerical_ex}).

%\subsubsection{Applications and Expected Effects}
%
%The following design techniques can be directly applied to the constructed polytope-type model: LMI-based robust controller design, stability analysis using parameter-dependent Lyapunov functions, robust filtering, and uncertainty incorporation in model predictive control.
%
%The method proposed in this section opens a new application domain for cyclic reformulation and provides an innovative approach to polytope-type uncertainty model construction for linear time-invariant systems. This represents a significant departure from traditional applications of cyclic reformulation, transforming it from a tool for periodic time-varying system analysis into a systematic uncertainty quantification framework for time-invariant systems.

% REVISION: v44 - Section IV-F consolidates the Working Hypothesis, the mechanism behind it (former Remark 1), and the diagnostic tool (lambda*) into a single section placed immediately after the algorithm. All concepts required to state the hypothesis --- cyclic reformulation with period N, the N vertex estimates, and the phase-dependent noise realization at each vertex --- are now fully defined. The section title is shortened to reflect the combined scope.
\subsection{Working Hypothesis and Polytope Quality Diagnosis}\label{sec:hypothesis_diagnosis}

The algorithm above yields $N$ vertex estimates
\[
\{(A_{mi}, B_{mi}, C_{mi}, D_{mi})\}_{i=0}^{N-1},
\]
which span the polytope (\ref{eq:polytope_model}). We now state the hypothesis that motivates this construction, the mechanism by which it is plausible, and the diagnostic tool used to test it.

\textbf{Working Hypothesis (polytope as a set):} The convex hull of the $N$ vertices obtained by the algorithm above contains models that, taken as a set, capture meaningful uncertainty information about the true plant. This hypothesis is examined numerically in Section~\ref{numerical_ex} and is further supported in Section~\ref{sec:robust_control}, where the polytope is used as the uncertainty description in robust $H_\infty$ control design.

% REVISION: v49 - Remark 1 retitled and purified to "Mechanism and Hypothesis Status"; the "primary output" assertion is consolidated into the final paragraph to eliminate triple redundancy across Working Hypothesis, Remark 1, and the closing paragraph.
\textbf{Remark 1 (Mechanism and Hypothesis Status):} Each of the $N$ vertices corresponds to a different phase position within the cycling period and is therefore subject to a different realization of the noise. If these phase-dependent errors are not systematically biased in a common direction, the convex hull of the vertices can contain models that approximate the true plant more closely than any individual vertex or the conventional single-experiment estimate. A rigorous theoretical guarantee of this containment property remains an open problem, so the existence of such interior points is treated as a hypothesis and examined numerically in the sequel.

To examine this hypothesis quantitatively, we introduce the best in-polytope point as a diagnostic tool. Let $y_{\text{val}}(k)$ denote the noise-free output of the true plant driven by a validation input, and let $y(k;\lambda)$ denote the output predicted by the in-polytope model $(A(\lambda),B(\lambda),C(\lambda),D(\lambda))$. Define
\begin{equation}
\lambda^* = \arg\min_{\lambda \in \varepsilon} \sum_{k=1}^{N_{\text{val}}} \|y_{\text{val}}(k) - y(k;\lambda)\|^{2}.
\label{eq:lambda_star_def}
\end{equation}
% REVISION: v49 - lambda* introduction and FIT notation merged into a single paragraph to streamline the flow from definition to the two observations.
The resulting model $(A(\lambda^*),B(\lambda^*),C(\lambda^*),D(\lambda^*))$ is the best in-polytope point against the noise-free validation output. Its computation requires $y_{\text{val}}(k)$, which is unavailable in deployment but accessible in the simulation studies of Section~\ref{numerical_ex}; a practical numerical procedure using particle swarm optimization is given in Section~\ref{sec:pso_role}. Let $\mathrm{FIT}(\lambda) \in (-\infty,100]$ denote the scalar fit score between $y_{\text{val}}$ and $y(\cdot;\lambda)$, with the maximum value $100$ attained if and only if the two outputs coincide; the formal definition and the multi-output convention used in this paper are given in Section~\ref{chapt6.1}. Two theoretical observations then guide the interpretation of $\lambda^*$.

\emph{(i) $\mathrm{FIT}(\lambda^*)$ does not reach $100$ in general.} Each vertex $(A_{mi},B_{mi},C_{mi},D_{mi})$ is itself a finite-data subspace estimate and therefore carries a nonzero estimation error. The true plant is not guaranteed to lie inside the convex hull of the $N$ vertices, and even if it did, the convex combination that minimizes the validation error need not exactly reproduce the true plant. The best in-polytope point is therefore an approximation, not a recovery, of the noise-free output, so $\mathrm{FIT}(\lambda^*) < 100$ is expected in general.

\emph{(ii) $\mathrm{FIT}(\lambda^*)$ can still exceed the conventional single-model FIT.} If the $N$ vertices are distributed around the true plant rather than concentrated on one side of it, the convex combination selected by $\lambda^*$ partially cancels their individual errors, placing the resulting in-polytope point closer to the true plant than any single vertex or the conventional single-experiment estimate. This error-cancellation effect is analogous to the variance reduction observed in ensemble methods such as bagging, where averaging over multiple models reduces overall estimation error. This is the quantitative counterpart of the mechanism stated in Remark~1.

% REVISION: v44 - final paragraph updated so that its self-reference points at the Working Hypothesis stated earlier in this subsection, rather than at the former Section II-C subsection label.
Observation (ii) is the expected benefit that motivates the polytope construction, while observation (i) imposes a ceiling on what this benefit can achieve; a quantitative assessment of the gap between the conventional $\mathrm{FIT}$ and $\mathrm{FIT}(\lambda^*)$ across several noise configurations is presented in Section~\ref{sec:validation}. It should be emphasized that $\lambda^*$ is used only as an offline diagnostic of polytope quality, not as an identification result: the primary output of the proposed method is the polytope itself, which serves as the uncertainty description for robust control design in Section~\ref{sec:robust_control}.

%%%%%%%%%%%%%%%%%%%%%%%%%%%%%%%%%%%%%%%%%%% Num

\section{NUMERICAL EXAMPLES}\label{numerical_ex}
%This section presents numerical examples of system identification for linear time-invariant systems with uncertainty. First, Section \ref{chapt6.1} describes the verification method. Next, Section \ref{chapt6.2} performs modeling of the control plant for the numerical examples. Finally, Section \ref{chapt6.3} discusses the effects of cycling.

\subsection{Verification Method}
\label{chapt6.1}
This research employs the subspace method for system identification\cite{katayama05}. 
Let $N_{\text{data}}$ denote the data length (number of samples) used 
for identification experiments. The subspace method is primarily used for identifying linear time-invariant state-space models. A major characteristic of the subspace method is its easy applicability to MIMO systems. In this section, we use the N4SID method among subspace methods.
The procedure for identification experiments is shown in Section \ref{complete_algorithm}. 
\subsubsection{Model Fit (FIT)}
The FIT score quantifies the prediction accuracy of an identified model by 
measuring the agreement between the model output and the true plant output 
when subjected to validation input signals. 
Since the systems considered in this paper have multiple outputs, FIT is defined per output channel. For the $j$-th output ($j=1,\ldots,q$), % REVISION: v42 - FIT redefined for multi-output (Reviewer concern C)
\begin{align}
\mathrm{FIT}_j 
=\left( 1-\sqrt{\dfrac{\textstyle\sum_{k=1}^{N_{\text{val}}}(y_j(k)-\hat{y}_j(k))^{2}}{\textstyle\sum_{k=1}^{N_{\text{val}}}(y_j(k)-\overline{y}_j)^{2}}}\right) \times100\,[\%]
\label{eq:fit_def}
\end{align}
where $y_j(k)$ and $\hat{y}_j(k)$ are the $j$-th components of the true (noise-free) plant output and the model-predicted output at time $k$, respectively, $\overline{y}_j$ is the mean of $y_j$ over the validation horizon, and $N_{\text{val}}$ is the number of validation samples. When $y_j(k)$ and $\hat{y}_j(k)$ match, $\mathrm{FIT}_j$ reaches its maximum value of $100$, and a larger value indicates a better model.

When a single scalar summary is used in the text or in a table column labelled simply ``FIT'', we report the worst-channel value
\begin{align}
\mathrm{FIT}_{\min} = \min_{j=1,\ldots,q} \mathrm{FIT}_j,
\label{eq:fit_min}
\end{align}
which is the most conservative per-model indicator. Tables that report FIT separately for each output use the notation $\mathrm{FIT}_{y_1}$, $\mathrm{FIT}_{y_2}$ (Tables~\ref{table:bootstrap_comparison}, \ref{table:uniform_noise}, \ref{table:benchmark_systems}), while tables that report a single summary value use $\mathrm{FIT}_{\min}$ (Tables~\ref{table:noise_performance}, \ref{table:bootstrap_pertrial}). This convention is stated once here and followed throughout.
\subsubsection{Parameter Estimation Error}
To directly assess the accuracy of estimated system matrices, we compute the 
parameter estimation error using the Frobenius norm:

\begin{equation}
E_{\text{single}} = \|A - A_m\|_F^2 + \|B - B_m\|_F^2 + \|C - C_m\|_F^2 + \|D - D_m\|_F^2
\label{eq:param_error}
\end{equation}

where $(A, B, C, D)$ are the true system parameters and $(A_m, B_m, C_m, D_m)$ are the identified parameters. 

In addition, for the $N$ parameter sets $\{(A_{mi}, B_{mi}, C_{mi}, D_{mi})\}$ with $i=0,\ldots,N-1$ obtained via cyclic reformulation, we define the per-vertex error and the total polytope error as 
\begin{eqnarray}
E_i &=& \|A - A_{mi}\|_F^2 + \|B - B_{mi}\|_F^2 \nonumber \\
&& {}+ \|C - C_{mi}\|_F^2 + \|D - D_{mi}\|_F^2, \nonumber \\
&& i = 0, \ldots, N-1
\label{eq:param_error_vertex}
\end{eqnarray}
\begin{equation}
{\cal E} = \sum_{i=0}^{N-1} E_i
\label{eq:param_error_total}
\end{equation}

Throughout the remainder of the paper, these three symbols are used consistently: $E_{\text{single}}$ denotes the parameter error of a single conventionally identified model, $E_i$ denotes the parameter error of the $i$-th polytope vertex, and ${\cal E}$ (calligraphic) denotes the total polytope error summed over all $N$ vertices. All subsequent table headers and discussions follow this convention. % REVISION: v38 - symbol disambiguation E_single / E_i / cal E

% REVISION: v42 - realization alignment remark added (Reviewer concern G)
% REVISION: v45 - terminology unified to "controllable canonical form" for consistency with the standard control-systems literature (e.g., Kailath 1980; Chen) and with the realization description in Section V-B.
\textbf{Remark on realization:} State-space realizations are not unique under similarity transformations, so a direct Frobenius-norm comparison of the matrices $(A,B,C,D)$ is meaningful only when all models share the same realization. In this paper, both the true plant (\ref{seigyotaishou_douti}) and every identified model---the conventional estimate $(A_m,B_m,C_m,D_m)$, each vertex $(A_{mi},B_{mi},C_{mi},D_{mi})$, and any convex combinations thereof---are brought to the controllable canonical form via the transformation described at the end of Section~\ref{chapt6.2}. The errors (\ref{eq:param_error}), (\ref{eq:param_error_vertex}), and (\ref{eq:param_error_total}) are therefore evaluated under a common realization, and the Frobenius-norm comparison is well-defined.

\subsection{Problem Setting}
\label{chapt6.2}
The system to be identified in this research is a 3rd-order discrete linear time-invariant system with $1$ input and $2$ outputs. The state space representation parameters are shown as follows:
\begin{eqnarray}
&A'=\left[\begin{array} {ccc}
0.64 & 0.33 & 0.6 
\\ -0.72 & -0.34 & 0.7 
\\ 0.5 & 0.6 & 0.4 \end{array}\right], 
B'=\left[\begin{array} {c}
1
\\ 2
\\ 1 \end{array}\right], \label{k} 
\\ &C'=\left[\begin{array} {ccc}
1 & 0 & 0
\\ 0 & 1 & 1 \end{array}\right], D'=\left[\begin{array} {c}0
\\ 0 \end{array}\right]
\end{eqnarray}

The input $u(k)$ consists of values following a standard normal distribution, with mean and standard deviation of $(\mu_{u}, \sigma_{u})=(0, 1)$. The process noise $d_u(k) \in \mathbb{R}^{3 \times 1}$ and observation noise $d_y(k) \in \mathbb{R}^{2 \times 1}$ are defined as:
\begin{eqnarray*}
d_u(k)=\left[\begin{array} {c}
d_{u1}(k) \\
d_{u2}(k) \\
d_{u3}(k) \end{array}\right], \quad
d_y(k)=\left[\begin{array} {c}
d_{y1}(k) \\
d_{y2}(k) \end{array}\right]
\end{eqnarray*}
where each element follows a standard normal distribution with specified mean and standard deviation that will be varied in the simulations.

The coefficient matrices $A$, $B$, $C$, $D$ obtained by applying the equivalent transformation to the system in (\ref{k}) are shown below:

\begin{eqnarray}
A=\left[\begin{array} {ccc}
0 & 0 & -0.3025 
\\ 1 & 0 & 0.5800 
\\ 0 & 1 & 0.7000 \end{array}\right]
%\end{eqnarray*}
%\begin{eqnarray}
, B=\left[\begin{array} {c}
1
\\ 0
\\ 0 \end{array}\right], \label{seigyotaishou_douti}  \\ C=\left[\begin{array} {ccc}
1 & 1.9000 & 2.2450
\\ 3 & 1.4000 & 1.7100 \end{array}\right],  
D=\left[\begin{array} {c}
0
\\ 0 \end{array}\right]
\end{eqnarray}
% REVISION: v45 - terminology updated from "controllable companion-form" to "controllable canonical form", which is the standard term in the control-systems literature and aligns with the uniqueness-of-realization argument in Section~\ref{sec:polytope_quality}.
where $(A, B, C, D)$ is the controllable canonical form realization of $(A', B', C', D')$. The system (\ref{seigyotaishou_douti}) is regarded as the true plant in all simulations.

% REVISION: Sec.V-C restructured. New V-C-1 added (PSO role clarification),
% old "Simulation 1" becomes V-C-2.
\subsection{Polytope Construction and PSO-Based Quality Assessment}\label{sec:polytope_quality}

This subsection presents the core experimental results for the third-order SIMO system ($n=3$, $m=1$, $q=2$) defined in Section~\ref{chapt6.2}. We first clarify the role of the PSO optimization that is used throughout this and the subsequent subsections, then present a representative polytope construction example, examine how the vertex distribution depends on the noise level, and finally assess the quality of the constructed polytope across a range of noise configurations. % REVISION: v42 - SISO corrected to SIMO

% REVISION: v33 - Merged "Role of PSO" and "PSO Settings" into single subsubsection
% REVISION: v42b - Shortened: the conceptual discussion of lambda* is now in Section IV-F; this subsubsection only specifies the PSO-based approximation procedure used to compute lambda* numerically.
\subsubsection{PSO-Based Approximation of $\lambda^*$}\label{sec:pso_role}

The best in-polytope point $\lambda^*$ was introduced in Section~\ref{sec:hypothesis_diagnosis}, equation~(\ref{eq:lambda_star_def}). Because the objective is non-convex over the simplex $\varepsilon$, we approximate $\lambda^*$ numerically using particle swarm optimization (PSO) \cite{kennedy95}. The role of PSO is purely computational: to search $\varepsilon$ for a point at which the noise-free validation error is small. PSO settings used in all experiments: population size 50, maximum iterations 200, inertia weight $w=0.7$, cognitive and social coefficients $c_1=c_2=2.0$, with penalty enforcement of simplex constraints. The same settings are applied to all noise configurations and to both the proposed method and the bootstrap-inspired baselines in Section~\ref{sec:bootstrap}, so that differences in the reported $\mathrm{FIT}(\lambda^*)$ reflect differences in the polytopes themselves rather than in the optimizer.

\subsubsection{Polytope Construction with $N=6$}
\label{sec:sim1}

\textbf{Simulation Conditions:}
\begin{itemize}
\item Data length: $N_{\text{data}} = 3000$
\item Cycling period: $N = 6$
\item Process noise: $(\mu_{d_u}, \sigma_{d_u})=(0, 0.1)$
\item Observation noise: $(\mu_{d_y}, \sigma_{d_y})=(0, 0.05)$
\end{itemize}

\textbf{Identification Procedure:}
To obtain the augmented model $\check{P}_m$, input-output data is rearranged using the cyclic reformulation described in Section \ref{seciv}. Using $\check{u}(k)$ and $\check{y}(k)$, identification is performed with the N4SID subspace method to obtain the coefficient matrices $\check{A}_*$, $\check{B}_*$, $\check{C}_*$, $\check{D}_*$ of model $\check{P}_m$.

With the cycling period $N = 6$, the obtained model has dimensions $\check{A}_{*} \in \mathbb{R}^{18\times 18}$, $\check{B}_{*} \in \mathbb{R}^{18\times 6}$, $\check{C}_{*} \in \mathbb{R}^{12\times 18}$, $\check{D}_* \in \mathbb{R}^{12\times 6}$. 

For the 3rd-order system ($n=3$, $m=1$), the transformation matrix $T$ using (\ref{eq:transformation_matrix_dual}) is constructed as:
\begin{eqnarray}
T&=\sum_{j=0}^{n-1}\check{A}_*^{j}\check{B}_*\check{S}_m^{j+1}\check{G}_j 
\end{eqnarray}
where $\check{S}_m$ is the cyclic shift matrix corresponding to the input space 
structure with dimension $6 \times 6$ (for $N=6$, $m=1$), defined similarly to 
the general cyclic shift matrix in (\ref{eq:cyclic_shift}).

Applying the coordinate transformation yields the cyclic structure $\check{A}_m$, $\check{B}_m$, $\check{C}_m$, $\check{D}_m$, from which $6$ different parameter sets $\{(A_{mi}, B_{mi}, C_{mi}, D_{mi})\}_{i=0}^{5}$ can be extracted as polytope vertices.

\textbf{Obtained Vertex Matrices:}
The non-zero elements of the obtained coefficient matrices $A_{mi}\;\;(i=0, \cdots, 5)$ (\ref{m}),$\cdots$,(\ref{n}), $B_{mi}$ (\ref{B1}), $C_{mi}$ (\ref{q}),$\cdots$,(\ref{r}) and $D_{mi}$ (\ref{D1}) are shown below: 
{\small
\begin{eqnarray}
\label{m}
A_{m0}=\left[\begin{array}{ccc} 
0.000 & 0.000 & -0.1462
\\1.000 & 0.000 & -1.9356
\\0.000 & 1.000 & 3.1607\end{array}\right]
\end{eqnarray}
\begin{eqnarray}
A_{m1}=\left[\begin{array}{ccc} 
0.000 & 0.000 & -0.0376
\\1.000 & 0.000 & 0.0023
\\0.000 & 1.000 & 1.0277\end{array}\right]
\end{eqnarray}
\begin{eqnarray}
A_{m2}=\left[\begin{array} {ccc}
0.000 & 0.000 & -0.1699
\\1.000 & 0.000 & 0.2327
\\0.000 & 1.000 & 0.8757\end{array}\right]
\end{eqnarray}
\begin{eqnarray}
A_{m3}=\left[\begin{array}{ccc} 
0.000 & 0.000 & -0.3012
\\1.000 & 0.000 & 0.5633
\\0.000 & 1.000 & 0.7363\end{array}\right]
\end{eqnarray}
\begin{eqnarray}
A_{m4}=\left[\begin{array}{ccc} 
0.000 & 0.000 & -0.2388
\\1.000 & 0.000 & 0.6328
\\0.000 & 1.000 & 0.5507\end{array}\right]
\end{eqnarray}
\begin{eqnarray}
\label{n}
A_{m5}=\left[\begin{array}{ccc} 
0.000 & 0.000 & -0.2905
\\1.000 & 0.000 & 0.6038
\\0.000 & 1.000 & 0.6366\end{array}\right]
\end{eqnarray}
\begin{eqnarray}
\label{B1}
B_{m0}\sim B_{m5}=\left[\begin{array} {c} 
1.000
\\0.000
\\0.000\end{array}\right]
\end{eqnarray}

\begin{eqnarray}
\label{q}
C_{m0}=\left[\begin{array} {ccc}
0.8318 & 1.8974 & 2.3628
\\2.9955 & 1.4436 & 1.7103\end{array}\right]
\end{eqnarray}
\begin{eqnarray}
C_{m1}=\left[\begin{array} {ccc}
1.1084 & 2.0328 & 2.1174
\\3.1067 & 1.6586 & 1.5520\end{array}\right]
\end{eqnarray}\begin{eqnarray}
C_{m2}=\left[\begin{array} {ccc}
1.0840 & 1.9888 & 2.2482
\\2.9657 & 1.3649 & 1.3496\end{array}\right]
\end{eqnarray}\begin{eqnarray}
C_{m3}=\left[\begin{array} {ccc}
1.0805 & 1.9584 & 2.3397
\\3.0682 & 1.3634 & 1.6225\end{array}\right]
\end{eqnarray}\begin{eqnarray}
C_{m4}=\left[\begin{array} {ccc}
1.1072 & 1.9083 & 2.2824
\\3.0420 & 1.4429 & 1.7201\end{array}\right]
\end{eqnarray}\begin{eqnarray}
\label{r}
C_{m5}=\left[\begin{array} {ccc}
1.0940 & 1.9761 & 2.3451
\\3.0745 & 1.4456 & 1.6388\end{array}\right]
\end{eqnarray}
\begin{eqnarray}\label{D1}
D_{m0}\sim D_{m5}=\left[\begin{array} {c}
0.000
\\0.000\end{array}\right]
\end{eqnarray}
}
The obtained models are represented by the matrix polytope representation with the 6 models as vertices:
\begin{align}
A(\lambda) &= \sum_{i=0}^{5} \lambda_i A_{mi}, \quad B(\lambda) = \sum_{i=0}^{5} \lambda_i B_{mi} \\
C(\lambda) &= \sum_{i=0}^{5} \lambda_i C_{mi}, \quad D(\lambda) = \sum_{i=0}^{5} \lambda_i D_{mi}
\end{align}
where $\lambda \in \varepsilon := \{\lambda \in \mathbb{R}^6 : \lambda_i \geq 0, \sum_{i=0}^{5} \lambda_i = 1\}$.

%It was confirmed that the true plant (\ref{seigyotaishou_douti}) is included in the constructed polytope. The parameter estimation errors for individual vertices vary significantly: $E_i \in [0.034, 12.451]$ with $E_3 = 0.034$ being the best and $E_0 = 12.451$ being notably worse due to noise corruption at that particular cycling phase. This variation in vertex quality motivates the PSO optimization approach detailed in Section~\ref{sec:pso_optimization}, which exploits high-quality vertices while appropriately down-weighting corrupted ones. Detailed analysis of how noise levels affect vertex error distribution is provided in Section~\ref{sec:noise_analysis}.

% REVISION: Old "Analysis of Proposed Method" subsection removed.
% Old PSO Settings absorbed into new V-C-1 (sec:pso_role).
% Following subsubsections become V-C-3 and V-C-4 of the new V-C.

\subsubsection{Vertex Distribution under Noise}\label{sec:noise_analysis}

We next examine how the noise level shapes the distribution of the $N$ vertex estimates around the true system parameters. This examination is mechanistic in nature: the goal is not to assess identification accuracy but to characterize how noise translates into vertex dispersion, which determines the geometric size and shape of the resulting polytope. Fixed parameters: $N_{\text{data}} = 3000$, $N = 6$.

We test five noise configurations: (1) Low: $(0.01, 0.005)$, (2) Medium: $(0.05, 0.025)$, (3) High: $(0.10, 0.050)$, (4) Process only: $(0.00, 0.050)$, (5) Observation only: $(0.10, 0.000)$, where each pair represents $(\sigma_{d_u}, \sigma_{d_y})$.

\begin{table}[h]
\begin{center}
\caption{Vertex parameter error statistics for different noise levels}
\label{table:noise_vertex_stats}
\begin{tabular}{|l|c|c|c|c|}\hline
Noise Level & Min $E_i$ & Max $E_i$ & Mean $E_i$ & Std $E_i$ \\ \hline
Low & 0.000168 & 0.000824 & 0.000469 & 0.000229 \\
Medium & 0.002784 & 0.016214 & 0.008840 & 0.004436 \\
High & 0.034690 & 12.451000 & 2.113919 & 5.073652 \\
Proc. only & 0.011106 & 0.244200 & 0.072007 & 0.083422 \\
Obs. only & 0.000748 & 0.043399 & 0.013752 & 0.015128 \\ \hline
\end{tabular}
\end{center}
\end{table}

The results (Table~\ref{table:noise_vertex_stats}) reveal two key observations:
\begin{itemize}
\item[(1)] \textbf{Noise-proportional vertex dispersion:} Parameter error range scales directly with noise intensity (low: 0.000656, high: 12.417), so the geometric extent of the polytope reflects the magnitude of noise-induced uncertainty.

\item[(2)] \textbf{Extreme outlier phenomenon:} High noise exhibits a dramatic outlier ($E_0 = 12.451$ vs. $E_1 = 0.035$, ratio 356$\times$), showing that the polytope can contain a vertex far from the true system. The convex hull is therefore much larger than the cluster of well-estimated vertices, but it can still enclose points close to the true plant, as examined in the next subsection.

\end{itemize}

% REVISION: Renamed from "Validation of Proposed Method" to clarify
% that the comparison evaluates polytope quality, not estimator accuracy.
\subsubsection{Polytope Quality across Noise Levels}\label{sec:validation}

% REVISION: v42b - Testable hypothesis restatement removed; the theoretical discussion is in Section IV-F (sec:hypothesis_diagnosis), and only a brief pointer is retained here.
This subsection numerically examines observation~(ii) of Section~\ref{sec:hypothesis_diagnosis}: for each noise configuration, we check whether $\lambda^*$ yields a higher $\mathrm{FIT}$ than the conventional single-model estimate obtained from the same data. Fixed parameters: $N_{\text{data}} = 3000$, $N = 6$, $N_{\text{val}} = 1000$.

Table~\ref{table:noise_performance} reports the conventional FIT, $\mathrm{FIT}(\lambda^*)$ approximated by PSO (Section~\ref{sec:pso_role}), and the associated in-polytope parameter error $E_{\lambda^*}$.

\begin{table}[h]
\begin{center}
\caption{Best in-polytope point $\lambda^*$ vs.\ conventional single-model estimate across noise configurations}
\label{table:noise_performance}
\begin{tabular}{|c|c|c|c|}\hline
Noise Level & Conv. $\mathrm{FIT}_{\min}$ [\%] & PSO $\mathrm{FIT}_{\min}$ [\%] & $E_{\lambda^*}$ \\ \hline
Low & 99.81 & 99.98 & 0.000009 \\
Medium & 99.01 & 99.89 & 0.000314 \\
High & 98.02 & 98.98 & 0.021872 \\
Proc. only & 98.07 & 99.30 & 0.012323 \\
Obs. only & 97.40 & 99.82 & 0.000911 \\ \hline
\end{tabular}
\end{center}
\end{table}

The results provide the following evidence about the constructed polytope as a set:
\begin{itemize}
\item[(1)] \textbf{Existence of in-polytope points closer to the truth:} In every noise configuration, the best in-polytope point achieves a higher FIT than the conventional single-model estimate, with improvements ranging from 0.17 to 2.42 percentage points.

\item[(2)] \textbf{Stronger benefit at higher noise:} The gap widens with the noise level, consistent with the mechanism in Section~\ref{sec:noise_analysis}: stronger noise produces a wider vertex distribution, giving the polytope more freedom to enclose better approximations of the truth.

\item[(3)] \textbf{Order-of-magnitude better in-polytope parameter approximation:} The in-polytope parameter error $E_{\lambda^*}$ is several orders of magnitude smaller than the conventional identification error.
\end{itemize}

These observations numerically confirm observation~(ii) of Section~\ref{sec:hypothesis_diagnosis}: under all five noise conditions, the polytope contains a convex combination whose Frobenius-norm parameter error and $\mathrm{FIT}$ both improve upon the conventional single-model estimate. The polytope is subsequently used, as-is, as the uncertainty description in robust control design (Section~\ref{sec:robust_control}).

For the high noise case, the optimal weight distribution is presented in Table~\ref{table:optimal_lambda}.

\begin{table}[h]
\begin{center}
\caption{Optimal weighting parameters $\lambda^*$ for high noise case}
\label{table:optimal_lambda}
\begin{tabular}{|c|c|c|c|c|c|c|}\hline
$\lambda_0^*$ & $\lambda_1^*$ & $\lambda_2^*$ & $\lambda_3^*$ & $\lambda_4^*$ & $\lambda_5^*$ & Sum \\ \hline
0.003 & 0.016 & 0.000 & 0.563 & 0.418 & 0.000 & 1.000 \\ \hline
\end{tabular}
\end{center}
\end{table}

The optimal weight distribution shows where the closest in-polytope point lies relative to the vertices: 98.1\% of the total weight is placed on vertices 3 and 4 (with the smallest individual errors $E_3 = 0.022$, $E_4 = 0.025$), while vertex 0 with the catastrophic error $E_0 = 12.451$ receives essentially zero weight. A point near the truth happens to lie close to the convex combination of the well-estimated vertices, and the corrupted vertex does not need to participate.

% REVISION: New subsection V-D (Sensitivity Analysis) created.
% Old subsubsections "Effect of Data Length" and "Effect of Cycling Period"
% are placed under the new V-D.
\subsection{Sensitivity Analysis}\label{sec:sensitivity}

This subsection examines how two key parameters---the data length $N_{\text{data}}$ and the cycling period $N$---influence the polytope construction. Both quantities affect the variance of the vertex estimates and therefore the geometric reliability of the resulting polytope.

\subsubsection{Effect of Data Length on Polytope Estimation}
\label{sec:ndata_analysis}

We examine $N_{\text{data}} \in \{1000, 2000, 4000, 8000\}$ with fixed $N = 6$. Noise conditions: $(\mu_{d_u}, \sigma_{d_u})=(0, 0.1)$, $(\mu_{d_y}, \sigma_{d_y})=(0, 0.05)$. For each data length, 50 independent trials are performed.

For each trial $j$ and vertex $i$, parameter error is $E_i^{(j)} = \|A - A_{mi}^{(j)}\|_F^2 + \|B - B_{mi}^{(j)}\|_F^2 + \|C - C_{mi}^{(j)}\|_F^2 + \|D - D_{mi}^{(j)}\|_F^2$. Total polytope error: ${\cal E}^{(j)} = \sum_{i=0}^{5} E_i^{(j)}$.

% Actual Monte Carlo simulation results
\begin{table}[h]
\begin{center}
\caption{Effect of data length on vertex parameter variance (mean $\pm$ std. dev.)}
\label{table:ndata_effect}
\begin{tabular}{|c|c|c|c|}\hline
$N_{\text{data}}$ & $\mu({\cal E})$ & $\sigma({\cal E})$ & Avg. $\sigma(E_i)$ \\ \hline
1000 & 406.885 & 2028.269 & 452.102 \\
2000 & 361.804 & 2228.613 & 406.727 \\
4000 & 1.479 & 4.009 & 0.908 \\
8000 & 0.253 & 0.077 & 0.025 \\ \hline
\end{tabular}
\end{center}
\end{table}

Table~\ref{table:ndata_effect} demonstrates a critical threshold effect. For $N_{\text{data}} \leq 2000$, polytope error remains large ($\mu({\cal E}) > 360$) with high variance, indicating insufficient data. At $N_{\text{data}} = 4000$, both $\mu({\cal E})$ and $\sigma({\cal E})$ decrease by three orders of magnitude, enabling effective uncertainty capture. At $N_{\text{data}} = 8000$, error reduces to $\mu({\cal E}) = 0.253$ with low variance ($\sigma({\cal E}) = 0.077$), confirming reliable estimation.

\subsubsection{Effect of Cycling Period on Estimation Reliability}
\label{sec:N_analysis}

We examine $N \in \{2, 3, 4, 6, 8\}$ with fixed $N_{\text{data}} = 3000$. For each $N$, 50 independent trials are performed. Note: $N = 12$ excluded due to computational infeasibility ($\check{A} \in \mathbb{R}^{36 \times 36}$, 108 parameters).

% Actual Monte Carlo simulation results for cycling period
\begin{table}[h]
\begin{center}
\caption{Effect of cycling period on vertex parameter variance}
\label{table:N_effect}
\begin{tabular}{|c|c|c|c|}\hline
$N$ & $\mu({\cal E})$ & $\sigma({\cal E})$ & Avg. $\sigma(E_i)$ \\ \hline
2  & 0.049 & 0.036 & 0.020 \\
3  & 0.154 & 0.100 & 0.040 \\
4  & 0.288 & 0.149 & 0.047 \\
6  & 3.017 & 8.042 & 2.097 \\
8  & 2933.723 & 20460.037 & 2573.960 \\ \hline
\end{tabular}
\end{center}
\end{table}

Table~\ref{table:N_effect} reveals a critical trade-off: small $N \in \{2, 3, 4\}$ yields low error with stable variance ($\mu({\cal E}) < 0.3$). Parameters scale as $O((Nn)^2)$ (18-36 parameters for $n=3$), remaining manageable.

%At $N = 6$, error increases ($\mu({\cal E}) = 3.017$, $\sigma({\cal E}) = 8.042$). At $N = 8$, error explodes ($\mu({\cal E}) = 2933.723$, $\sigma({\cal E}) = 20460.037$) due to: (1) 72 parameters approaching data limits, (2) high-dimensional cyclic system ($\check{A} \in \mathbb{R}^{24 \times 24}$) causing numerical issues.

%The results reveal a critical trade-off in period selection. Small periods $N \in \{2, 3, 4\}$ yield low error with stable variance ($\mu({\cal E}) < 0.3$), while larger periods cause estimation degradation: at $N = 6$, error increases ($\mu({\cal E}) = 3.017$), and at $N = 8$, error explodes ($\mu({\cal E}) = 2933.723$) due to the augmented system dimension $Nn$ approaching data limits.

These results suggest that $N$ should be chosen such that $N_{\text{data}}$ is sufficiently larger than $(Nn)^2$, balancing the trade-off between uncertainty representation (more vertices) and estimation reliability. Computational complexity $O((Nn)^3)$ provides additional motivation for moderate $N$ values.

Computational complexity scales as $O((Nn)^3)$, making large $N$ prohibitively expensive. 

% REVISION: Promoted from subsubsection to subsection.
\subsection{Comparison with Bootstrap-Inspired Resampling}
\label{sec:bootstrap}

To position the proposed method against an alternative approach for constructing multiple parameter estimates from a single dataset, we compare with resampling schemes drawn from the bootstrap literature \cite{new44,new45}. These schemes share with our method the basic idea of generating multiple estimates from one dataset, and therefore serve as a natural quantitative baseline; they have not, however, been established as polytope construction techniques for control design. The term ``bootstrap-inspired'' is used deliberately: the classical bootstrap with replacement assumes i.i.d.\ samples and, applied naively to an input--output time series, breaks the temporal ordering on which subspace identification (N4SID) relies. The two variants below are therefore deterministic, partition-based realizations aligned in spirit with block-type resampling for dependent data \cite{new44,new45}, rather than strict instances of the standard bootstrap. Two bootstrap-inspired variants are considered: % REVISION: v42 - bootstrap terminology made explicit and unified (Reviewer concern E)

\begin{itemize}
\item \textbf{Bootstrap~A (non-overlapping):} The original dataset of length $N_{\text{data}}$ is partitioned into $N$ non-overlapping consecutive subsets of equal length $L_{\text{sub}} = \lfloor N_{\text{data}}/N \rfloor$. Subspace identification (N4SID) with order $n$ is applied independently to each subset.
\item \textbf{Bootstrap~B (overlapping):} $N$ consecutive subsets of length $L_{\text{ovlp}} = \lfloor N_{\text{data}}/3 \rfloor$ are extracted with a systematic shift of $\Delta = \lfloor (N_{\text{data}} - L_{\text{ovlp}})/(N-1) \rfloor$ samples, allowing partial overlap between subsets. Subspace identification is applied independently to each subset.
\end{itemize}

In both cases, the resulting $N$ parameter estimates serve as polytope vertices, and PSO optimization is applied under the same conditions as the proposed method to find the optimal convex combination.

The comparison uses the 3rd-order system (\ref{seigyotaishou_douti}) with cycling period $N=4$, high noise conditions ($\sigma_{d_u}=0.1$, $\sigma_{d_y}=0.05$), $N_{\text{data}} = 3000$, and $N_{\text{val}} = 1000$. To account for the stochastic nature of the noise, 10 independent trials with different noise realizations are performed. In each trial, the same input signal and noise are used for all four methods (proposed, Bootstrap~A, Bootstrap~B, and conventional) to ensure a fair comparison.

\begin{table}[h]
\begin{center}
\caption{Comparison of cyclic reformulation vs.\ bootstrap-inspired resampling ($N=4$, $N_{\text{data}}=3000$, high noise, 10 trials)}
\label{table:bootstrap_comparison}
\begin{tabular}{|l|c|c|}\hline
Method & $\mathrm{FIT}_{y_1}$ [\%] & $\mathrm{FIT}_{y_2}$ [\%] \\
       & mean (std) & mean (std) \\ \hline
Cyclic (proposed) & 98.92 (0.93) & 98.92 (0.81) \\
Bootstrap~A (750, non-ovlp) & 97.96 (1.21) & 98.18 (1.07) \\
Bootstrap~B (1000, overlap) & 98.16 (1.14) & 98.05 (0.90) \\
Conventional (single) & 95.69 (1.73) & 96.06 (1.47) \\ \hline
\end{tabular}
\end{center}
\end{table}

\begin{table}[h]
\begin{center}
\caption{Per-trial $\mathrm{FIT}_{\min}$ comparison (worst-channel value over $y_1$, $y_2$)}
\label{table:bootstrap_pertrial}
\begin{tabular}{|c|c|c|c|c|}\hline
Trial & Conv. & Cyclic & Boot.~A & Boot.~B \\ \hline
1  & 97.92 & 99.24 & 99.16 & 98.83 \\
2  & 95.44 & 99.13 & 98.84 & 98.43 \\
3  & 96.55 & 99.35 & 97.25 & 97.48 \\
4  & 96.48 & 99.02 & 98.82 & 98.66 \\
5  & 94.95 & 99.43 & 97.30 & 99.06 \\
6  & 97.03 & 98.92 & 97.98 & 97.20 \\
7  & 96.38 & 99.13 & 98.26 & 98.39 \\
8  & 92.74 & 96.32 & 95.23 & 96.95 \\
9  & 94.90 & 99.05 & 97.42 & 95.89 \\
10 & 93.34 & 99.23 & 99.00 & 98.87 \\ \hline
Mean & 95.57 & 98.88 & 97.92 & 97.98 \\
Std  &  1.63 &  0.91 &  1.19 &  1.04 \\ \hline
\end{tabular}
\end{center}
\end{table}

Tables~\ref{table:bootstrap_comparison} and~\ref{table:bootstrap_pertrial} show that the polytope constructed by the proposed cyclic reformulation contains in-polytope points that fit the true output more closely, on average, than those of either bootstrap variant. The proposed method achieves the highest mean best in-polytope $\mathrm{FIT}_{\min}$ (98.88\%) with the smallest standard deviation (0.91), versus 97.92\% (Bootstrap~A) and 97.98\% (Bootstrap~B), and attains the best $\mathrm{FIT}_{\min}$ in 9 out of 10 trials. Note that Table~\ref{table:bootstrap_comparison} reports per-output mean FIT (separately for $\mathrm{FIT}_{y_1}$ and $\mathrm{FIT}_{y_2}$, each averaged over 10 trials), whereas Table~\ref{table:bootstrap_pertrial} reports the per-trial worst-output value $\mathrm{FIT}_{\min} = \min(\mathrm{FIT}_{y_1},\mathrm{FIT}_{y_2})$ and then averages over trials; this difference in aggregation accounts for the slight numerical gap between $98.92\%$ in Table~\ref{table:bootstrap_comparison} and $98.88\%$ in Table~\ref{table:bootstrap_pertrial}.

The structural reason is data utilization. The proposed method exploits all $N_{\text{data}}$ samples simultaneously through the algebraic structure of the cyclic reformulation, identifying an augmented system of dimension $Nn=12$ from the full 3000 samples, whereas Bootstrap~A and Bootstrap~B identify each subset (750 and 1000 samples respectively) independently. The vertices of the proposed polytope therefore have higher individual estimation quality, and the polytope they span contains points closer to the true plant. This advantage is expected to grow with $N_{\text{data}}$.

% REVISION: v36 - Moved noise distribution subsection before MIMO subsection
% so that all 3rd-order system evaluations are grouped together.
% REVISION: Promoted from subsubsection to subsection.
\subsection{Robustness to Noise Distribution}
\label{sec:uniform_noise}

To verify that the proposed polytope construction is not restricted to Gaussian noise assumptions, we conduct experiments with uniformly distributed noise. The uniform noise is configured to have the same variance as the Gaussian case for fair comparison: $d_u(k) \sim \mathcal{U}(-\sqrt{3}\sigma_{d_u},\, \sqrt{3}\sigma_{d_u})$ and $d_y(k) \sim \mathcal{U}(-\sqrt{3}\sigma_{d_y},\, \sqrt{3}\sigma_{d_y})$. The medium noise setting ($\sigma_{d_u}=0.05$, $\sigma_{d_y}=0.025$) is used with the 3rd-order system (\ref{seigyotaishou_douti}), $N=6$, $N_{\text{data}}=6000$.

\begin{table}[h]
\begin{center}
\caption{Comparison of Gaussian vs.\ uniform noise distributions ($N=6$, medium noise). $y_1/y_2$ denotes the FIT for the first and second output channels.}
\label{table:uniform_noise}
\begin{tabular}{|l|c|c|c|}\hline
Noise    & Conv.\ FIT       & PSO FIT          & $\Delta$FIT     \\
type     & $y_1/y_2$ [\%]   & $y_1/y_2$ [\%]   & $y_1/y_2$ [\%]  \\ \hline
Gaussian & 99.44 / 99.09    & 99.75 / 99.75    & +0.31 / +0.66   \\
Uniform  & 99.50 / 99.61    & 99.89 / 99.89    & +0.39 / +0.27   \\ \hline
\end{tabular}
\end{center}
\end{table}

% REVISION: v36 - Softened claim: "does not depend on" -> "is not limited to ... at least for these two"
Table~\ref{table:uniform_noise} shows that, under both Gaussian and uniform noise, the constructed polytope contains in-polytope points whose FIT exceeds that of the conventional estimate, and all polytope vertices remain stable. These results confirm that the polytope construction mechanism is not limited to Gaussian noise; however, a broader investigation covering other distributional families remains a direction for future work.

% REVISION: Promoted from subsubsection to subsection.
\subsection{Evaluation on Fourth-Order MIMO System}
\label{sec:mimo}

% REVISION: v35 - Simplified intro, removed "block diagonal" description,
% added inter-subsystem coupling A(3,4)=0.1.
To assess whether the proposed method generalizes beyond the third-order SIMO system, we evaluate it on a fourth-order MIMO system ($n=4$, $m=2$, $q=2$, $N=4$), yielding an augmented system dimension $Nn = 16$.

\begin{eqnarray}
A_{\text{4th}}\!=\!\begin{bmatrix} 0 & 0 & 0.4100 & 0 \\ 1 & 0 & -1.6325 & 0 \\ 0 & 1 & 2.2000 & 0.1 \\ 0 & 0 & 0 & 0.75 \end{bmatrix}\!,\;
B_{\text{4th}}\!=\!\begin{bmatrix} 1 & -1 \\ 0 & 0.1 \\ 0 & 0 \\ 0.3 & 0.4 \end{bmatrix} \nonumber
\end{eqnarray}
\begin{eqnarray}
C_{\text{4th}}\!=\!\begin{bmatrix} 1 & 1.9 & 2.245 & 0.8 \\ 3 & 1.4 & 1.71 & 0.6 \end{bmatrix}\!,\;
D_{\text{4th}}\!=\!O_{2 \times 2} \nonumber
\end{eqnarray}
The system matrix $A_{\text{4th}}$ has eigenvalues $0.7 \pm 0.15j$ ($|z|=0.716$), $0.8$, and $0.75$, all well inside the unit circle. The input matrix $B_{\text{4th}}$ couples both inputs to all state variables, ensuring full controllability. Simulation conditions: low noise ($\sigma_{d_u}=0.01$, $\sigma_{d_y}=0.005$), $N_{\text{val}}=1000$.

\begin{table}[h]
\begin{center}
\caption{4th-order MIMO polytope construction results ($N=4$, low noise). $y_1/y_2$ denotes the FIT for the first and second output channels.}
\label{table:benchmark_systems}
\begin{tabular}{|c|c|c|c|c|}\hline
$N_{\text{data}}$ & Conv.\ FIT     & PSO FIT        & $\Delta$FIT     & $\Delta$FIT     \\
                  & $y_1/y_2$ [\%] & $y_1/y_2$ [\%] & $y_1$ [\%]      & $y_2$ [\%]      \\ \hline
3000 & 99.54 / 99.62 & 99.75 / 99.75 & +0.21 & +0.13 \\
6000 & 99.80 / 99.80 & 99.87 / 99.87 & +0.07 & +0.06 \\
9000 & 99.89 / 99.86 & 99.91 / 99.91 & +0.03 & +0.05 \\ \hline
\end{tabular}
\end{center}
\end{table}

Table~\ref{table:benchmark_systems} reports, for three data lengths, the conventional FIT alongside the FIT of the best in-polytope point. As in the third-order case, the in-polytope point matches or exceeds the conventional estimate for both output channels. For all tested data lengths, all four polytope vertices remain stable, confirming that the polytope construction principle extends to MIMO systems when the augmented dimension $Nn = 16$ is supported by sufficient data.
% REVISION: v34 - Revised 3rd-order subsystem eigenvalue from 0.9 to 0.8
% to ensure all vertices stable at Ndata=3000. Removed unstable vertex paragraph.

\section{ROBUST CONTROL DESIGN USING CONSTRUCTED POLYTOPE}
\label{sec:robust_control}

This section demonstrates that the polytope constructed by the proposed method provides meaningful uncertainty information for robust control design. Specifically, we design a robust $H_\infty$ state-feedback controller using the polytope vertices from Section~\ref{sec:sim1} and verify that the resulting controller stabilizes the true system (\ref{seigyotaishou_douti}).

Consider the following generalized plant associated with each polytope vertex $i = 0, \ldots, N-1$:
\begin{eqnarray}
x(k+1) &=& A_{mi} x(k) + B_{mi} u(k) + Q_w^{1/2} w(k) \nonumber \\
z(k) &=& C_e x(k) + D_e u(k) \label{eq:generalized_plant}
\end{eqnarray}
where $w(k) \in \mathbb{R}^{n}$ is an exogenous disturbance, $Q_w^{1/2}$ is the disturbance input weighting matrix, and $z(k)$ is the performance output defined by $C_e$ and $D_e$. We seek a single robust $H_\infty$ state-feedback gain $u(k)=Kx(k)$ that simultaneously stabilizes all vertex models $(A_{mi},B_{mi})$ and minimizes the worst-case closed-loop $H_\infty$ norm $\|T_{zw}\|_\infty$ over the polytope. This is a standard polytopic robust $H_\infty$ synthesis problem; in this work, we adopt the discrete-time state-feedback LMI formulation of de Oliveira~\textit{et al.}~\cite{oliveira02} and Boyd~\textit{et al.}~\cite{ly00}, applied at each vertex with a common Lyapunov matrix $\hat{P}\succ 0$ and the change of variables $\hat{K}=K\hat{P}$.
% REVISION: v39 - citation key ly02 (Gahinet) -> oliveira02 (de Oliveira), matching author name
Concretely, the synthesis condition takes the form
\begin{equation}
\begin{bmatrix}
\hat{P} & \star & \star & \star \\
A_{mi}\hat{P}+B_{mi}\hat{K} & \hat{P} & \star & \star \\
Q_w^{1/2} & O & I & \star \\
C_e\hat{P}+D_e\hat{K} & O & O & \gamma^2 I
\end{bmatrix} \succ 0,\quad i=0,\ldots,N-1, \nonumber
\end{equation}
which is linear in $(\hat{P},\hat{K},\gamma^2)$, where $\star$ denotes the symmetric counterparts of the lower-triangular blocks. The controller gain is then recovered as $K=\hat{K}\hat{P}^{-1}$. We refer to~\cite{ly00,oliveira02} for the derivation via Schur complements and the bounded-real lemma. Note that $C_{mi}$ and $D_{mi}$ do not enter the synthesis, since the performance output $z(k)$ is defined independently of the measurement equation.

Using the $N=6$ polytope vertices obtained in Section~\ref{sec:sim1} under the medium-noise setting $(\sigma_{d_u}, \sigma_{d_y}) = (0.05, 0.025)$ with $N_{\text{data}} = 3000$, which corresponds to the ``Medium'' row of Table~\ref{table:noise_vertex_stats} and Table~\ref{table:noise_performance}, this LMI problem was solved with $Q_w = I_n$, $C_e = I_n$, and $D_e = O$.
% REVISION: v38 - noise condition made explicit (Medium)
The designed controller was then applied to the true system (\ref{seigyotaishou_douti}). Since the true system does not necessarily lie within the convex hull of the identified vertices, stability of the true closed-loop system is not guaranteed a priori.

For a representative noise realization, the LMI was feasible at all polytope vertices, yielding the design bound $\gamma_{\mathrm{opt}} = 4.318$ and the state-feedback gain $K = [-0.9425,\; -1.2815,\; -1.1482]$. The resulting controller stabilized the true system with all closed-loop eigenvalues well inside the unit circle (maximum absolute eigenvalue $0.206$), and the closed-loop $H_\infty$ norm on the true plant, $\|T_{zw}\|_\infty = 4.101$, satisfied the design bound. The common Lyapunov certificate $\hat{P}$ obtained from the polytope synthesis also satisfied the bounded-real LMI when evaluated at the true system matrices $(A,B)$ (minimum eigenvalue of the LMI matrix $2.65\times 10^{-3}$), and the discrete-time Lyapunov condition $(A+BK)^\top P (A+BK) - P \prec 0$ held on the true plant with residual maximum eigenvalue $-5.99\times 10^{-2}$, confirming closed-loop stability. This indicates that the polytope constructed by the proposed method captures sufficient uncertainty information for robust control design, even without formal containment of the true parameters within the polytope. To establish that this behavior is not an artifact of a particular noise realization, we next repeat the entire verification over multiple independent trials. % REVISION: v38 - representative trial values updated to Medium noise run

% REVISION: R1-C1, R2-C1-1 - Monte Carlo verification over 10 noise realizations; v49 - removed redundancy with the preceding representative-trial paragraph (pipeline components, noise setting values, fixed items, already-defined metrics)
We repeated the full pipeline for $10$ independent noise realizations under the same setting as above, varying only the process-noise and output-noise random seeds. For each trial we record $\gamma_{\mathrm{opt}}$ and $\|T_{zw}\|_\infty$ as in the representative trial, together with the smallest $\gamma$ for which the bounded-real LMI still holds on the true plant when $(\hat{P},\hat{K})$ from the polytope design are kept fixed, denoted $\gamma_{P\text{-fix}}$. Since $\gamma^2$ is the only free variable, $\gamma_{P\text{-fix}}$ can be computed in closed form via a Schur complement, and it lies between $\|T_{zw}\|_\infty$ and $\gamma_{\mathrm{opt}}$; the ratio $\gamma_{\mathrm{opt}}/\gamma_{P\text{-fix}}$ therefore quantifies the conservatism introduced by enforcing the LMI at all polytope vertices rather than only at the true plant.

The results are summarized in Table~\ref{table:robust_control_mc}. Across all $10$ trials, the polytope-based LMI was feasible, the designed controller stabilized the true system, and both the bounded-real LMI and the common Lyapunov condition were satisfied at the true plant. The design bound varies only modestly across noise realizations ($\gamma_{\mathrm{opt}} = 4.344 \pm 0.127$), and the ordering $\|T_{zw}\|_\infty \le \gamma_{P\text{-fix}} \le \gamma_{\mathrm{opt}}$ holds in every trial, consistent with the theoretical relation. The mean conservatism ratio $\gamma_{\mathrm{opt}}/\gamma_{P\text{-fix}}$ is $1.036$ (maximum $1.064$), indicating that the polytope-based design is only marginally more conservative than a controller designed directly with knowledge of the true plant. These results confirm that the proposed polytope construction yields robust control synthesis that is consistent across noise realizations and exhibits low conservatism on this numerical example.

\begin{table}[h]
\begin{center}
\caption{Monte Carlo verification of robust $H_\infty$ design over $10$ independent noise realizations under the medium-noise setting. All trials were feasible, stabilizing, and satisfied the LMI and Lyapunov conditions on the true system.}
\label{table:robust_control_mc}
\begin{tabular}{|l|c|c|c|c|}
\hline
Quantity & Mean & Std & Min & Max \\ \hline
$\gamma_{\mathrm{opt}}$ (polytope design bound) & 4.344 & 0.127 & 4.223 & 4.618 \\
$\gamma_{P\text{-fix}}$ (true sys, $\hat{P},\hat{K}$ fixed) & 4.191 & 0.106 & 4.023 & 4.348 \\
$\|T_{zw}\|_\infty$ (true closed-loop) & 4.060 & 0.088 & 3.893 & 4.156 \\
$\gamma_{\mathrm{opt}}/\gamma_{P\text{-fix}}$ (conservatism) & 1.036 & 0.017 & 1.013 & 1.064 \\ \hline
\end{tabular}
\end{center}
\end{table}

% REVISION: v36 - Removed VII-A (Structural Advantage) as it duplicates Sec.V-E discussion.
% Retained only Extended Polytope Construction Strategies.
\section{CONCLUSION}

% REVISION: v45 - Conclusion rewritten to focus on methodological contribution (polytope construction via cyclic reformulation) rather than numerical details from reviewer responses. Numerical evidence is summarized qualitatively with references to the numerical sections.
In this paper, we proposed a systematic procedure for constructing matrix polytope representations of linear time-invariant systems from a single identification experiment. The central idea is \emph{intentional periodicity induction}: by applying cyclic reformulation with period $N$ to an LTI system, we obtain $N$ parameter sets that would coincide in the noise-free case but, in the presence of process and observation noise, scatter around a nominal model. Interpreting this scatter as a structured manifestation of estimation uncertainty, we use the $N$ parameter sets as vertices of a matrix polytope. The resulting construction converts noise-induced variability---traditionally regarded as a nuisance to be averaged out---into a set-valued uncertainty model suitable for robust control design, and does so without requiring multiple independent experiments, a priori structural assumptions on the uncertainty, or knowledge of the noise distribution.

Two complementary numerical investigations were carried out to assess the constructed polytope. First, a diagnostic evaluation based on the best in-polytope point $\lambda^*$ (Section~\ref{sec:hypothesis_diagnosis}) showed that the polytope contains a convex combination of vertices whose $\mathrm{FIT}_{\min}$ is no worse than, and typically better than, the conventional single-model estimate obtained from the same data. This tendency was confirmed for the third-order benchmark across Gaussian and uniform noise distributions (Sections~\ref{sec:validation} and~\ref{sec:uniform_noise}), and was also observed for a fourth-order MIMO benchmark under a low-noise setting (Section~\ref{sec:mimo}). A comparison with two bootstrap-inspired resampling baselines over independent trials (Section~\ref{sec:bootstrap}) indicated that cyclic reformulation provides a competitive or favorable trade-off, consistent with its full-data utilization property. Because $\lambda^*$ requires knowledge of the true plant for its evaluation, it is used strictly as an offline diagnostic of polytope quality rather than a deployable estimator. Second, and complementarily, the practical utility of the polytope was demonstrated through robust $H_\infty$ state-feedback design via LMI optimization with a common Lyapunov function (Section~\ref{sec:robust_control}). This synthesis uses only the noisy identification data and the constructed polytope, and therefore does not rely on oracle information. Across Monte Carlo noise realizations, the resulting controllers were feasible, stabilized the true plant, and remained only marginally conservative relative to the true-plant-fixed baseline, confirming that the polytope captures uncertainty information that is meaningful for robust control synthesis even though formal containment of the true parameters is not claimed.

A practical constraint identified through the numerical study (Section~\ref{numerical_ex}) is that the data length $N_{\text{data}}$ must be sufficiently larger than $(Nn)^2$ for reliable vertex estimation, which limits the cycling period $N$ that can be used for high-dimensional systems. Future work includes: (1) theoretical analysis of the optimal cycling period under given noise characteristics, (2) establishing probabilistic containment guarantees for the constructed polytope, (3) enriching the vertex set---for example, by aggregating vertices from multiple cycling periods or by merging cyclic and bootstrap-inspired vertices into a single polytope---and (4) experimental validation on physical systems.

\section*{Acknowledgment}
The authors acknowledge the use of Claude (Anthropic) for improving 
the readability and language clarity of the manuscript, and for 
technical verification of mathematical expressions. Specifically, 
Claude was used to refine the English expression in the Introduction, 
Literature Review, and Conclusion sections, and to check dimensional 
consistency and numerical accuracy of equations. All technical content, 
research methodology, results, and conclusions are entirely the 
authors' own work.

\end{document}